\documentclass[10pt]{amsart}
\usepackage[utf8]{inputenc}
\usepackage{amsmath}
\usepackage{amssymb}
\usepackage{graphicx}
\usepackage{algorithm,algorithmic}
\usepackage{appendix}

\newtheorem{de}{Definition}
\newtheorem{theo}{Theorem}
\newtheorem{prop}{Proposition}

\newtheorem{cor}{Corollary}
\renewcommand{\bigwedge}{\Lambda}
\newcommand{\Diff}{\text{Diff}}

\title[Functional currents and functional shapes]{Functional currents : a new mathematical tool to model and analyse functional shapes}
\author{Nicolas Charon, Alain Trouv\'{e}}

\begin{document}

\maketitle

\begin{abstract}
This paper introduces the concept of \emph{functional current} as a mathematical framework to represent and treat functional shapes, i.e. sub-manifold supported signals. It is motivated by the growing occurrence, in medical imaging and computational anatomy, of what can be described as \emph{geometrico-functional} data, that is a data structure that involves a deformable shape (roughly a finite dimensional sub manifold) together with a function defined on this shape taking value in another manifold. 

Indeed, if mathematical currents have already proved to be very efficient theoretically and numerically to model and process shapes as curves or surfaces (\cite{Durrlemann} and \cite{Glaunes}), they are limited to the manipulation of purely geometrical objects. We show that the introduction of the concept of functional currents offers a genuine solution to the simultaneous processing of the geometric and signal information of any functional shape. We explain how functional currents can be equipped with a Hilbertian norm mixing geometrical and functional content of functional shapes nicely behaving under geometrical and functional perturbations and paving the way to various processing algorithms. We illustrate this potential on two problems: the redundancy reduction of functional shapes representations through matching pursuit schemes on functional currents and the simultaneous geometric and functional registration of functional shapes under diffeomorphic transport.

\end{abstract}
\tableofcontents
\newpage
\section{Introduction}
\label{sec:intro}
 Shape analysis is certainly one the most challenging problem in pattern recognition and computer vision \cite{Cootes2001,Kendall1999,Dryden1998,Belongie2002,Michor2003}. Moreover, during the last decade, shape analysis has played a major role in medical imaging through the emergence of computational anatomy \cite{Grenander1998,Thompson2002,Miller2002,Ashburner2007,Miller2009,Pennec2011}. More specifically, the quest of anatomical biomarkers through the analysis of normal and abnormal geometrical variability of anatomical manifolds has fostered the development of innovative mathematical frameworks for the representation and the comparison of a large variety of geometrical objects. Among them, since their very first significant emergence in the field of computational anatomy, \emph{mathematical currents} have become more and more commonly used framework to represent and analyse shapes of very various natures, from unlabelled landmarks to curves (\cite{Glaunes}), fiber bundles (\cite{Durrlemann4}) surfaces (\cite{Glaunes2}) or 3D volumes. The reasons of this success, which we shall detail in the next section, lie basically in the generality of the framework with respect to a very wide collection of geometrical features as well as in their robustness to change of topology and of parametrization. The crucial step at this point is to define a proper distance between currents that faithfully transcribes variations of geometry itself. This problem has been successfully addressed by embedding current spaces into Reproducing Kernel Hilbert Spaces (RKHS), providing kernel-based norms on currents which are fully geometric (independent of parametrization) and enable practical computations in a very nice setting. Such norms and the resulting distances allow to define attachment terms between the geometrical objects, which are then used for instance to drive registration algorithms on shapes (\cite{Glaunes2}, \cite{Glaunes}) and perform statistical analysis of their variability (\cite{Durrlemann4}, \cite{Durrlemann3}). 

More recently though, an increasing number of data structures have emerged in computational anatomy that not only involve a geometrical shape but some signal attached on this shape, to which we give the general name of \textit{functional shapes}. The most basic example is, of course, classical images for which the geometrical support is simply a rectangle on which is given a 'grey level' signal. In many cases however, the support can have a much more complex geometry like, for instance, the activation maps on surfaces of cortex obtained through fMRI scans. Signals can also include structures that are more sophisticated than simple real values : we could think of a vector field on a surface as well as tensor-valued signal that appear in DTI imaging. Such a diversity both in shape and signal makes it a particularly delicate issue to embed all geometrico-functional objects in one common framework. Despite several attempts to model them directly as currents, important limitations of currents were found in such problems, which we will develop in section 2. As a result, recent approaches have been rather investigating methods where shape and signal are treated separately instead of trying to define an attachment distance between geometrico-functional objects. This is the case for instance in \cite{Sabuncu} where authors propose a registration algorithm for fMRI data in which is performed an anatomic matching followed by a second one based on the values of the signals. However, all these frameworks have two important drawbacks : they are first very specific to a certain type of dataset and they require an exact one to one correspondence between the two shapes in order to further compare functional values, whereas in many applications inexact matchings are far more appropriate. 

The purpose of this paper is to \emph{describe} and \emph{explore} a new analytical setting to work on the most general problem of \emph{representation} and \emph{comparison} of geometrico-functional structures (compatible with any change of parametrisation of their geometrical supports) treated as elements of an embedding functional vector space, here a Reproducing Kernel Hilbert Space, on which many desirable operations can be performed. 

Our new analytical setting shares some common features with the \emph{mathematical current} setting that will be recalled briefly in section \ref{sec:cur} but overcome its main limitations when dealing with functional shapes. The core idea, developed in section \ref{sec:def} is to \emph{augment} usual currents with an extra component embedding the signal values by a natural tensor product leading to our definition of
\textit{functional currents}. We consider then various actions on functional currents by diffeomorphic transport in section \ref{sec:def} and shows in section \ref{sec:hilbert} that kernel norms can provide a suitable Hilbertian structure on functional currents generalizing greatly what has been done for currents. We also show in what sense this representation and RKHS metric on functional currents is consistent with the idea of comparing functional shapes with respect to deformations between them, which makes it a good approach for defining attachment distances. The two main results on this topic are the control results of propositions \ref{propcontrolnorm} and \ref{propcontrolnormdefor}. We then illustrate the potential of this new metric setting in section \ref{sec:processing} on two different problems. The first illustration is the construction, via a matching pursuit algorithm, of redundancy reduction or compression algorithm of the representation of functional shapes by functional currents with few examples of compression on curves and surfaces with real-valued data. The second illustration is about the potential benefits of functional currents in the field of computational anatomy. In particular, we show a few basic results of diffeomorphic matching between functional shapes with our extension of large deformation diffeomorphic metric mapping (LDDMM) algorithm \cite{Beg2005} to functional currents.

\section{Currents in the modelling of shapes}
\label{sec:cur}
\subsection{A brief presentation of currents in computational anatomy}
Currents were historically introduced as a generalization of distributions by L. Schwartz and then G. De Rham in \cite{DeRham}. The theory was later on considerably developed and connected to geometric measure theory in great part by H. Federer \cite{Federer}. In the first place, these results found interesting applications in calculus of variations as well as differential equations. However, the use of currents in the field of computational anatomy is fairly more recent since it was considered for the first time in \cite{Glaunes}. In the following, we try to outline the minimum background of theory about currents needed to recall the link between shapes and currents. 

First of all, we fix some notations. Let's call $E$ a generic euclidean space of dimension $n$. We will denote by $\Omega_{0}^{p}(E)$ the space of continuous $p$-differential forms on $E$ that vanish at infinity. Every element $\omega$ of $\Omega_{0}^{p}(E)$ is then a continuous function such that for all $x \in E$, $\omega(x) \in \bigwedge^{p} E^{*}$. Since we have the isomorphism $\bigwedge^{p} E^{*} \approx \left ( \bigwedge^{p} E \right )^{*}$, we can see both $\omega(x)$ as a p-multilinear and alternated form on $E$ and as a linear form on the $\binom{n}{p}$-dimensional space of $p$-vectors in $E$. For all the following, we will use the notation $\omega_{x}(\xi)$ as the evaluation of a differential form $\omega$ at point $x \in E$ and on the $p$-vector $\xi$. On $\bigwedge^{p} E$ can be defined an euclidean structure induced by the one of $E$, which is such that if $\xi=\xi_{1}\wedge..\wedge\xi_{p}$ and $\eta=\eta_{1}\wedge..\wedge\eta_{p}$ are two simple $p$-vectors, $\langle \xi, \eta \rangle= \text{det}(\langle \xi_{i}, \eta_{j}\rangle)_{i,j}$. The norm of a simple $p$-vector is therefore the volume of the element. The space $\Omega_{0}^{p}(E)$ is then equipped with the infinite norm of bounded functions defined on $E$. These notations adopted, we define the space of $p$-currents on $E$ as the topological dual $\Omega_{0}^{p}(E)'$, i.e. the space of linear and continuous forms on $\Omega_{0}^{p}(E)$. Note that in the special case where $p=0$, the previous definition is exactly the one of usual distributions on $E$ that can be also seen as signed measures on $E$. Simplest examples of currents are given by generalization of a Dirac mass : if $x\in E$ and $\xi \in \bigwedge^{p}E$, $\delta_{x}^{\xi}$ is the current that associates to any $\omega \in \Omega_{0}^{p}(E)$ its evaluation $\omega_{x}(\xi)$.

Now, the relationship between shapes and currents lies fundamentally in the fact that \textit{every d-dimensional and oriented sub-manifold $X$ of $E$ of finite volume can be represented by an element of $\Omega_{0}^{p}(E)'$}. Indeed, we know from integration theory on manifolds (\cite{Federer},\cite{Lang}) that any d-differential form of $\Omega_{0}^{p}(E)$ can be integrated along $X$, which associates to $X$ a $d$-current $C_{X}$ such that :
\begin{equation}
\label{sbman_current}
 C_{X}(\omega)=\int_{X} \omega
\end{equation}
for all $\omega \in \Omega_{0}^{p}(E)$. The application $X\mapsto C_{X}$ is also injective. Equation \eqref{sbman_current} can be rewritten in a more explicit way if $X$ admits a parametrization given by a certain smooth immersion $F:U \rightarrow E$ with $U$ an open subset of $\mathbb{R}^{d}$. Then,
\begin{equation*}
 C_{X}(\omega)=\int_{(x_{1},..,x_{d}) \in U} \omega_{F(x_{1},..,x_{d})} \left ( \frac{\partial F}{\partial x_{1}} \wedge ... \wedge \frac{\partial F}{\partial x_{d}} \right ) dx_{1}...dx_{d}\,. 
\end{equation*}
It is a straightforward computation to check that the last expression is actually independent of the parametrization (as far as the orientation is conserved). In the general case, there always exists a partition of the unit adapted to the local charts of $X$, so that $C_{X}$ could be expressed as a combination of such terms. \textit{The representation is fully geometric in the sense that it only depends on the manifold structure itself and not on the choice of a parametrization}. Currents' approach therefore allows to consider sub-manifolds of given dimension (curves, surfaces,...) as elements of a fixed functional vector space. This also gives a very flexible setting to manipulate shapes since addition, combination or averages become straightforward to define. On the other hand, spaces of currents contain a lot more than sub-manifolds because general currents do not usually derive from sub-manifolds (think for instance of a punctual current $\delta_{x}^{\xi}$). However, it encompasses in a unified approach a wide variety of geometrical objects as for instance sets of curves and surfaces which can be relevant in some anatomy problems.

In registration issues, a fundamental operation is the transport of objects by a diffeomorphism of the ambient space. If $C \in \Omega_{0}^{p}(E)'$ and $\phi \in \Diff(E)$, we define the transport of $C$ by $\phi$ as the classical push-forward operation denoted $\phi_{\sharp} C$ :
\begin{equation}
\label{push_courants}
 \forall \omega \in \Omega_{0}^{p}(E), \ \left ( \phi_{\ast} C \right )(\omega)= C \left ( \phi^{\ast}\omega \right )
\end{equation}
where $\phi^{\ast}\omega$ is the usual pull-back of a differential form defined for all $x \in E$ and $\xi=\xi_{1}\wedge...\wedge \xi_{p} \in \bigwedge^p E$ by :
\begin{equation}
\label{pull_formes}
 \left ( \phi^{\ast}\omega \right )_{x}(\xi)=\omega_{\phi(x)}(d_{x}\phi(\xi_{1})\wedge...\wedge d_{x}\phi(\xi_{p}))
\end{equation}
$d_{x}\phi$ being the notation we use for the differential of the diffeomorphism at point $x$. With this definition, it's a straightforward proof to check that $\phi_{\ast} C_{X}=C_{\phi(X)}$, which means that \textit{the $d$-current associated to a submanifold transported by $\phi$ is the $d$-current associated to the transported submanifold $\phi(X)$}.

To complete this brief presentation of currents applied to computational anatomy, we still need to explain how the currents' representation can be practically implemented and how computations can be made on them. This step consists mainly in approximating the integral in \eqref{sbman_current} into a discrete sum of punctual currents $ C_{X} \approx \sum_{k=1..N} \delta_{x_{k}}^{\xi_{k}}$ where $x_{k}$ are points in E and $d$-vectors $\xi_{k}$ encode local elements of volume of the manifold $X$. A manifold $X$ would be then stored as a list of $N$ \textit{momenta} $\delta_{x_{k}}^{\xi_{k}}$ consisting of points' coordinates and corresponding $d$-vectors. However, the transition between $X$ and its approximation as a discrete current cannot usually be performed in a standard way. Computationally, a mesh on the sub-manifold is needed. Let's examine the two most frequent cases of curves and surfaces. Let $\gamma : I \rightarrow E$ be a continuous curve in E given by a sampling of $N$ points $\{x_{k}=\gamma(t_{k})\}_{k=1..N}$. Starting from this approximation of $\gamma$ as a polygonal line, we can associate the $1$-current defined by :
\begin{equation*}
 \tilde{C}_{\gamma}=\sum_{j=1}^{N-1} \delta_{c_{j}}^{\tau_{j}}
\end{equation*}
with $c_{j}$ the center of segment $[x_{j} x_{j+1}]$ and $\tau_{j}$ the vector $x_{j+1}-x_{j}$. 
It can be proved easily that $|C_{X}(\omega)-\tilde{C}_{X}(\omega)| $ tends toward zero for all $1$-form $\omega$ as $ \max_{k} \{|t_{k+1}-t_{k}|\}\rightarrow 0 $, i.e. as the sampling gets more accurate (cf \cite{Glaunes}). Same process can be applied to a triangulated surface S immersed in $E=\mathbb{R}^{3}$. We associate to each triangle of the mesh $x_{j}x_{j+1}x_{j+2}$ a punctual current $\delta_{c_{j}}^{\xi_{j}}$ with $c_{j}=\frac{1}{3}(x_{j}+x_{j+1}+x_{j+2})$ and $\xi_{j}=\frac{1}{2} (x_{j+1}-x_{j})\wedge (x_{j+2}-x_{j})$. Since we have $\bigwedge^{2}\mathbb{R}^{3} \approx \mathbb{R}^{3}$, the previous formal $2$-vector can be identified to the usual wedge product of vectors in $\mathbb{R}^{3}$, that is the normal vector to the surface whose norm encodes the area of the triangle. Again, it can be shown that this approximated current gets closer and closer to the actual $C_{S}$ as the mesh is refined. Eventually, the surface is represented as a finite collection of points and normal vectors in the space E. 

Finally, the question of building a metric on the space of currents should be addressed. There are several norms traditionally defined on $\Omega_{0}^{p}(E)'$ such as the mass norm or the flat norm. However, those are either not easily computable in practice or unfitted to comparison between shapes (see \cite{Durrlemann} chap 1.5). A particularly nice framework to avoid both problems is to define a Hilbert space structure on currents through reproducing kernel Hilbert space (RKHS) theory. This approach consists in defining a vector kernel on E ($K:E \times E \rightarrow \mathcal{L}(\bigwedge^{p}E)$) and its associated RKHS $W$. Under some assumptions on the kernel, it can be shown that the space of $p$-currents is continuously embedded in the dual $W'$ which is also a Hilbert space. Therefore, in applications, we generally consider $W'$ instead of $\Omega_{0}^{p}(E)'$ as our actual space of currents. For more details on the construction of RKHS on currents, we refer to \cite{Durrlemann} and \cite{Glaunes}. Since, in applications, manifold are represented by sums of punctual currents, it's sufficient to be able to compute inner products between two punctual currents. RKHS framework precisely gives simple closed expressions of such products. Indeed, one can show that $\langle \delta_{x_{1}}^{\xi_{1}}, \delta_{x_{2}}^{\xi_{2}}\rangle_{W'}=\xi_{1}^{T}K(x_{1},x_{2})\xi_{2}$. Computation of distances between shapes then reduces to simple kernel calculus which can be performed efficiently for well-suited kernels either through fast Gauss transform schemes as in \cite{Glaunes} or through convolutions on linearly spaced grids as explained in \cite{Durrlemann}. 

In summary, this succinct presentation was meant to stress two essential advantages of currents in shape representation. The first one being its flexibility due to the vector space structure and the wide range of geometrical objects that are comprehended without ever requiring any parametrization. The second important point is the fact that computations on currents are made very efficient by the use of kernels which makes them appropriate in various applications as simplification, registration or template estimation. All these elements motivate an extension of the framework of currents to incorporate functional shapes, which will be discussed thoroughly in all the following.

\begin{figure}
\leftskip -1cm
\begin{tabular}{cc}
\includegraphics[width=6cm,height=4cm]{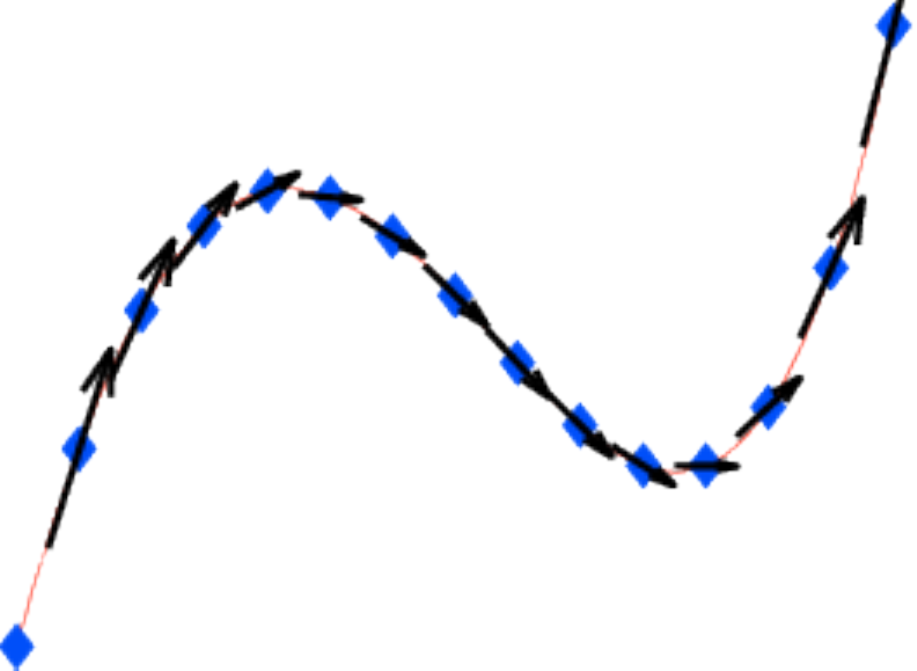}
\includegraphics[width=7cm,height=4.5cm]{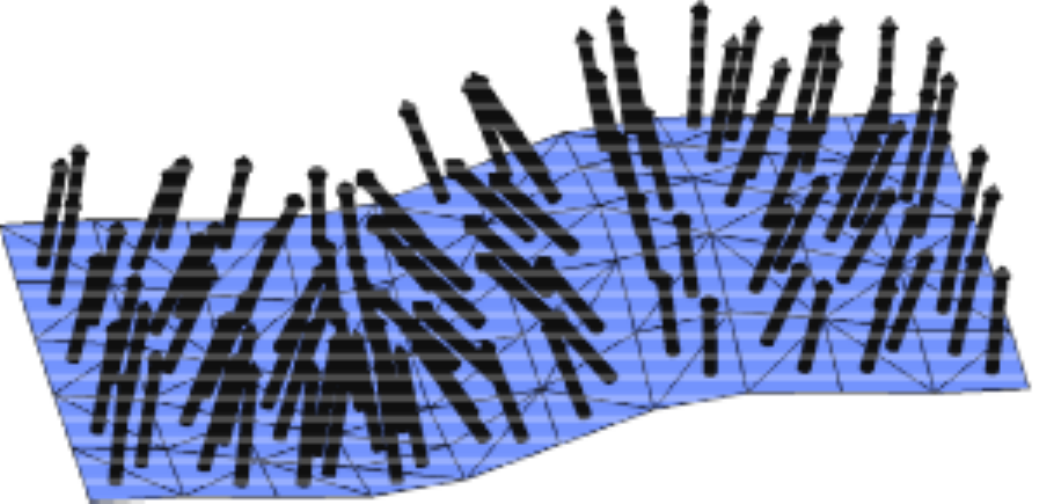}
\end{tabular}
\caption{Representation of curve and surface in Dirac current}\label{discretisations_courants}
\end{figure}

\subsection{Functional shapes and the limitations of currents}
\label{subsec:func}
We now consider, as in the previous section, a $d$-dimensional sub-manifold $X$ of the $n$-dimensional vector space $E$ but in addition, we assume that functional data is attached to every points of $X$ through a function $f$ defined on $X$ and taking its values in a differentiable manifold $M$, the \textit{signal space}. What we call a \textbf{functional shape} is then a couple $(X,f)$ of such objects. The natural question that arises is this : can we model such functional shapes in the framework of currents like purely geometrical shapes~? In the following, we are discussing two possible methods to address this question directly with usual currents and explain why both of them are not fully satisfying in the perspective of applications to computational anatomy. 

\begin{figure}[htb]
\leftskip -1cm
\begin{tabular}{cc}
\includegraphics[width=7cm]{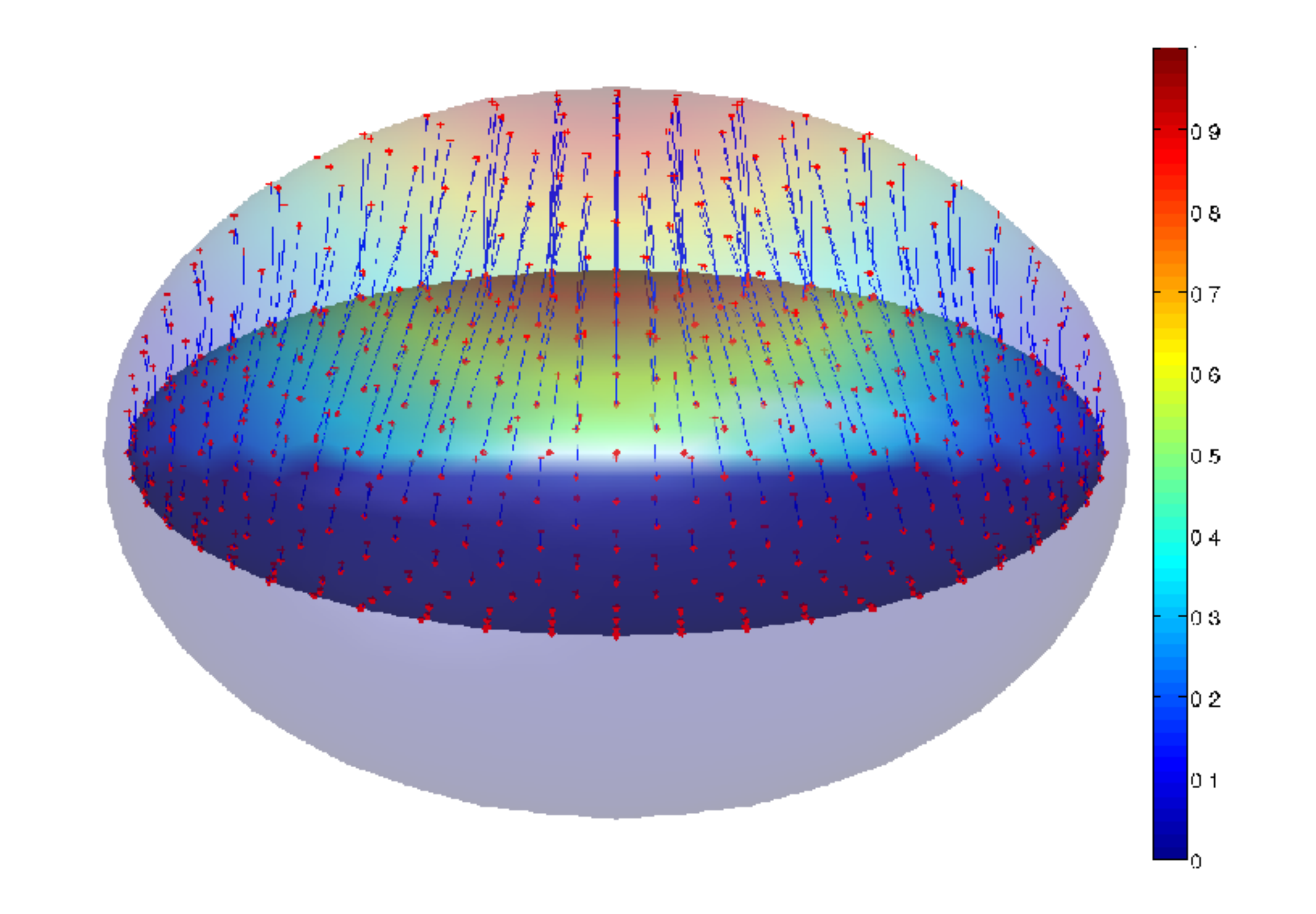}
\includegraphics[width=7cm]{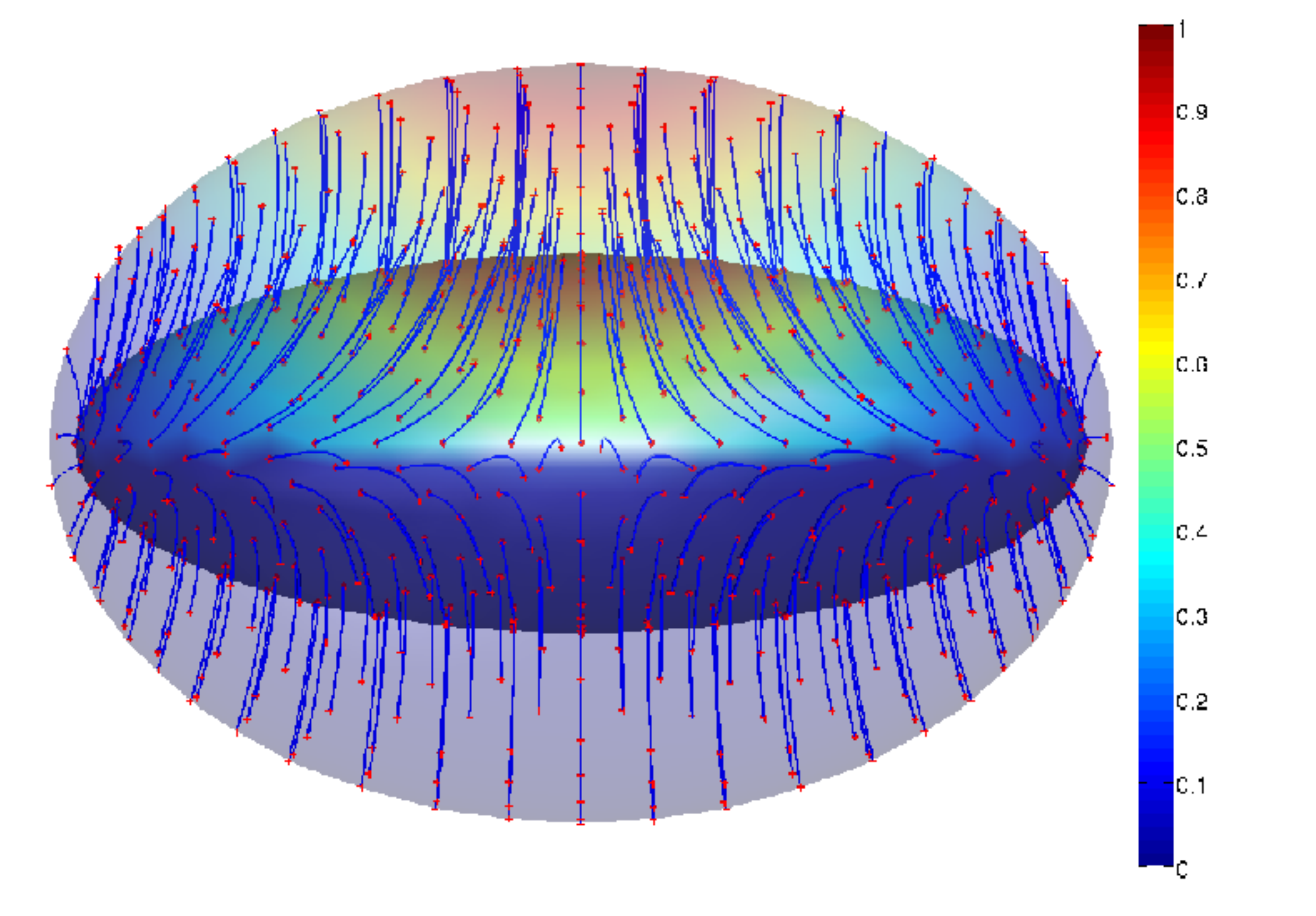}
\end{tabular}
\caption{An example of matching between two ellipsoids provided by the classical LDDMM algorithm. On the left, the adaptation with the colored currents' representation. Values of the signals are two diffused stains both on the source ellipsoid (inside surface) and the target one (exterior shaded surface). We display in blue trajectories of the points. The points compounding to zero-valued area of the signal in the source shape are not matched to the corresponding points in the target surface. On the right, we show what should be the expected result. It is obtained through the approach of functional currents that shall be presented in the next parts of the paper.}\label{courantscolor}
\end{figure}

First attempts to include signals supported geometrically in the currents' representation were investigated in \cite{Durrlemann} with the idea of \textit{colored currents}. This relies basically on the fact already mentioned that the set of $d$-currents contains a wider variety of objects than $d$-dimensional sub-manifolds like rectifiable sets or flat chains (cf \cite{Federer}). In particular, weighted sub-manifolds can be considered as currents in the following very natural way : suppose that $X$ is a sub-manifold of $E$ of dimension $d$ and $f : \ X \rightarrow \mathbb{R}$ is a weight or equivalently a real signal at each point of $X$ such that $f$ is continuous, then we can associate to $(X,f)$ a $d$-current in E :  
\begin{equation*}
 T_{(X,f)}(\omega) = \int_{X} f\omega
\end{equation*}
Although this approach seems to be the most straightforward way to apply currents to functional shapes since we are still defining a $d$-current in $E$, it's quite obvious that such a representation suffers from several important drawbacks. The first thing is the difficulty to generalize colored currents for signals that are not simply real-valued, particularly if the signal space is not a vector space (think for instance of the case of a signal consisting of directions in the 3D space, where $M$ is therefore the sphere $\mathbb{S}^{2}$). 
The second point arises when the previous equation is discretized into Dirac currents, which leads to an expression of the form $\sum_{k=1..N} f(x_{k})\delta_{x_{k}}^{\xi_{k}}$. We notice an ambiguity appearing between the signal and the volume element $\xi$ since for any $r \neq 0$, $f(x_{k})\delta_{x_{k}}^{\xi_{k}}=r f(x_{k})\delta_{x_{k}}^{\xi_{k}/r}$ ; separating geometry from signal in the discretized version appears as a fundamental difficulty. In addition, the energy of Dirac terms are proportional to the value of the signal at the corresponding point which induces an asymmetry between low and high-valued signals. In this setting, areas having very small signals become negligible in terms of current, which is both not justified in general and can affect drastically the matching of colored currents. We show a simple illustration of this issue when matching two colored ellipsoids with this approach in figure \ref{courantscolor}.  Finally, we could also mention some additional pitfalls resulting in that colored currents do not separate clearly geometry from signal. Most problematic is the fact that there is no flexibility to treat signals at different scale levels than geometry which can make the approach highly sensitive to noise.

\begin{figure}[!h]
\begin{tabular}{cc}
   \includegraphics[width=6cm]{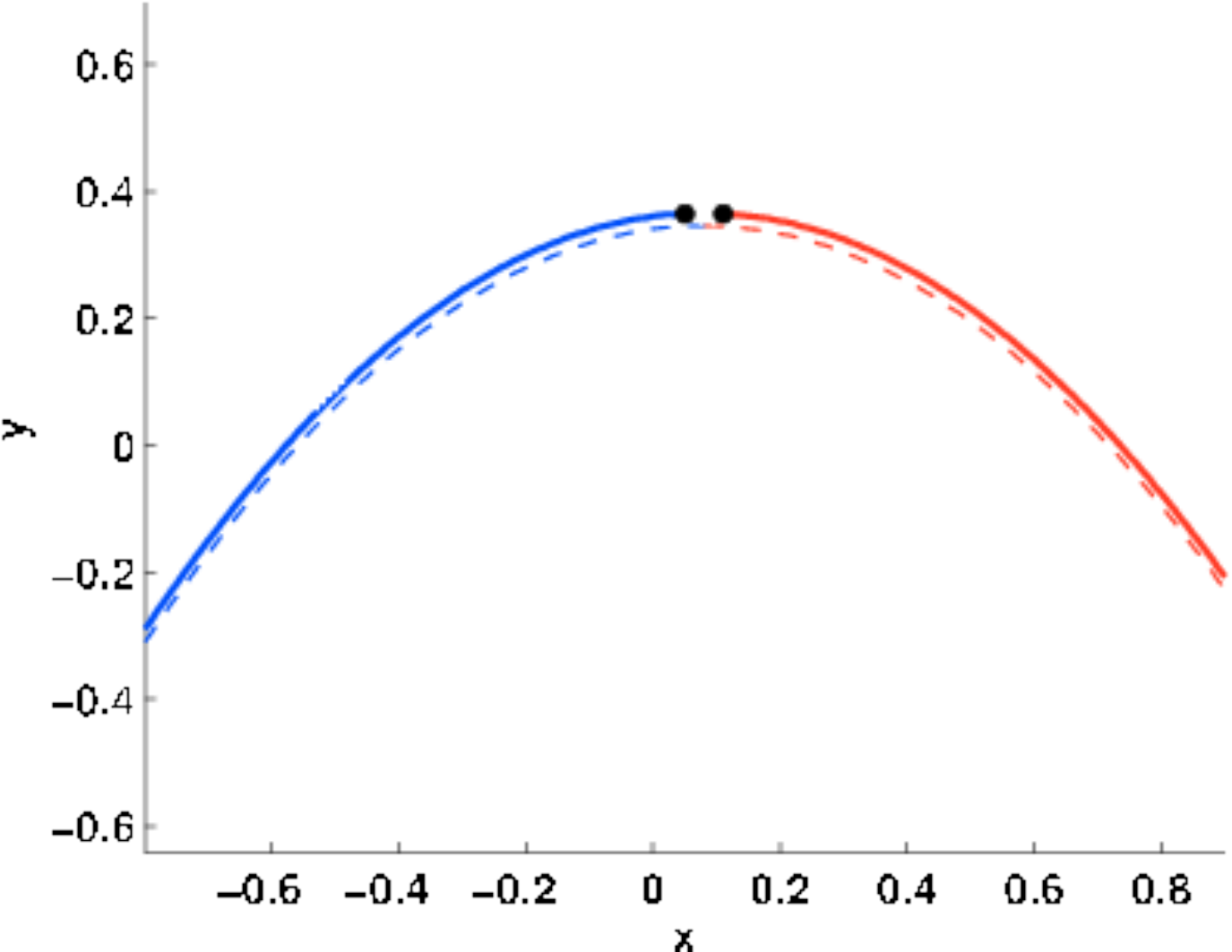} &
   \includegraphics[width=7cm]{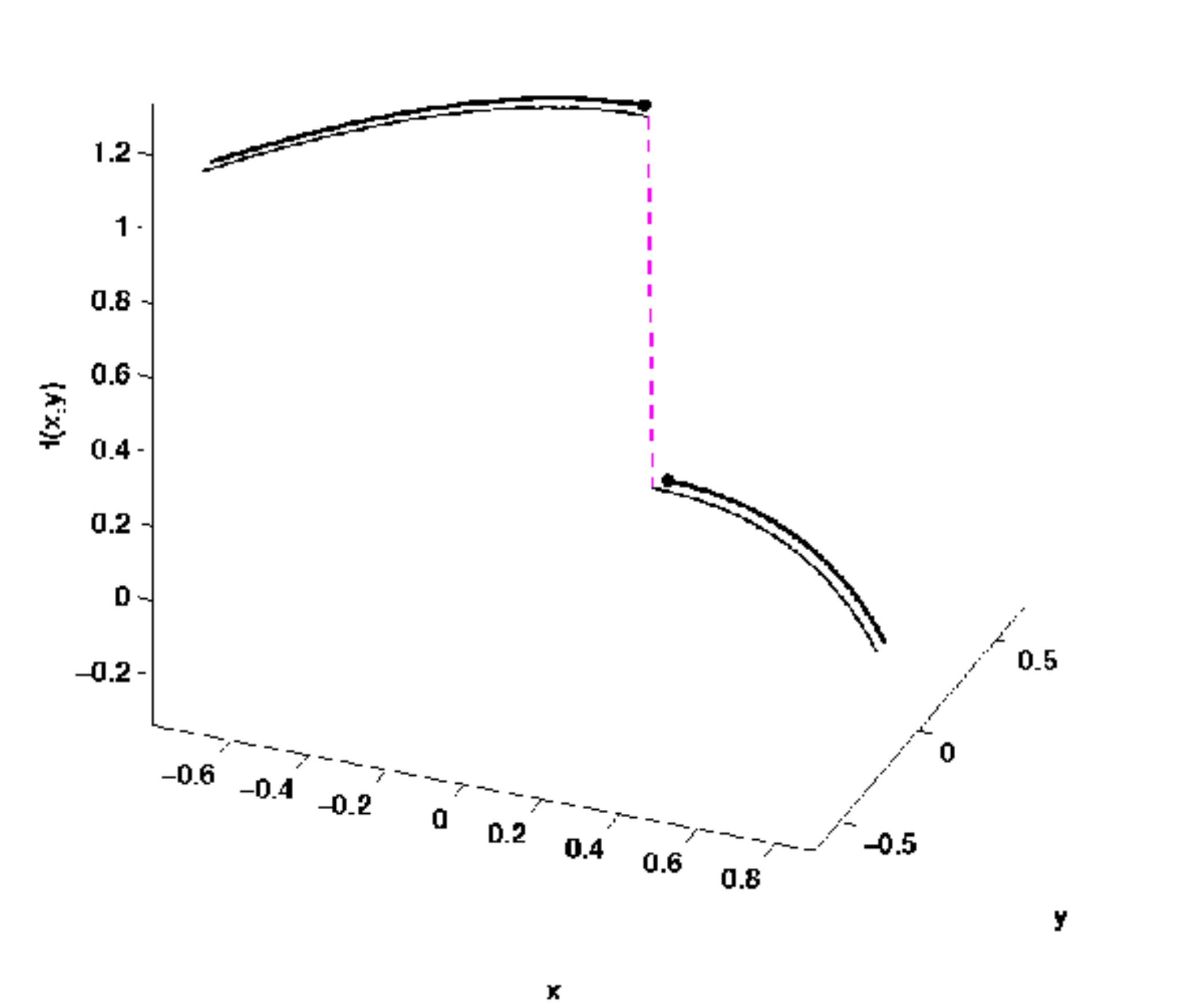} \\
\end{tabular}
\caption{Product currents and topology. On the left, we show a disconnected 2D curve with signal values $0$ in blue and $1$ in red as well as the connected curve in dashed line. On the right hand side are the corresponding curves in the 3-dimensional geometry $\times$ signal space. What we want to emphasize here is the fact that no RKHS norm on product currents would provide a continuity of this representation with respect to connectivity : the difference between the two curves is the magenta dashed part which represents a pure variation in the signal domain.} \label{courants_prod}
\end{figure}

Another possible and interesting way to represent a functional shape by a current is to view it as a shape in the product space $E \times M$. Somehow, it generalizes the idea of seeing a 2D image as a 3D surface. However, at our level of generality, it is not a completely straightforward process. If the signal function $f$ is assumed to be $C^{1}$, the set $G:=\{(p,f(p))\ |\  p \in X\}$ inherits a structure of $d$-dimensional manifold of $E \times M$. With $M$ a vector space, it results directly from the previous that $G$ can be represented as a $d$-current in the product space, that is as an element of $\Omega_{0}^{d}(E \times M)$. For a general signal manifold though, we would need to extend our definitions of currents to the manifold case, which could be done (cf \cite{DeRham}) but the definition of kernels on such spaces would then become a much more involved issue in general compared to the vector space case. This difficulty set apart, there still are some important elements to point out. The first one is the increase of dimensionality of the approach because, while we are still considering manifold of dimension $d$, the co-dimension is higher : the space of $d$-vectors characterizing local geometry $\bigwedge^{d}(E \times M)$ is now of dimension $\binom{n+dim(M)}{d}$, with significant consequences from a computational point of view. From a more theoretical angle, we see that, in such an approach, geometrical support and signal play a symmetric role. In this representation, the modelled topology is no more the one of the original shape because we also take into account variations within the signal space. Wether this is a strength or a weakness is not obvious a priori and would highly depend on the kind of applications. What we can state is that this representation is not robust to topological changes of the shape : in practice, the connectivity between all points becomes crucial, what we illustrate on the simplest example of a plane curve carrying a real signal in figure \ref{courants_prod}. In the field of computational anatomy, the processing of data such as fiber bundles, where connections between points of the fibers are not always reliable, this would be a clear drawback. We shall illustrate these consequences from the point of view of diffeomorphic matching in the last section of the paper.   \\
To sum up this section, we have investigated two direct ways to see a functional shape as a current. The colored currents' setting, although being very close to the modelling of purely geometrical shapes, is to be discarded mainly because it mixes geometry and signal in an inconsistent way. As for the second idea of immersing the functional shape in a product space, we have explained its limits both from the difficulty of the practical implementation and from the lack of robustness with respect to topology of the geometrical support. These facts constitute our motivation to redefine a proper class of mathematical objects that would preserve the interest of currents while overcoming the previous drawbacks.

\section{Definition and basic properties of functional currents}
\label{sec:def}
In this section, we propose an extension of the notion of currents to represent functional shapes. The new mathematical objects we introduce, we call 'functional currents', are \textbf{not} usual currents strictly speaking, contrarily to the methods presented in section \ref{subsec:func}. They would rather derive from the very general concept of double current introduced originally by De Rham in \cite{DeRham}. Here, we adapt it in a different way to fit with the applications we aim at in computational anatomy.

\subsection{Functional $p$-forms and functional currents}
Like in the previous section, let $(X,f)$ be a functional shape, with $X$ a $d$-dimensional sub-manifold of the $n$-dimensional Euclidean space $E$ and $f$ a measurable application from $X$ to a signal space $M$. In our framework, $M$ can be any Riemannian manifold. Most simple examples are provided by surfaces with real signal data like activation maps on cortex in fMRI imaging but the framework that we present here is made general enough to incorporate signals from very different natures : vector fields, tensor fields, grassmannians. We now define the space of functional currents again as the dual of a space of continuous forms :
\begin{de}
 We call a functional $p$-form on $(E,M)$ an element of the space $C_{0}(E\times M,\bigwedge^{p}E^{\ast})$ which will be denoted by $\Omega_{0}^{p}(E,M)$ hereafter. We consider the uniform norm on $\Omega_{0}^{p}(E,M)$ defined by : $\|\omega\|_{\infty}=\sup_{(x,m)\in E \times M} \|\omega_{(x,m)}\|$.  A functional $p$-current (or \emph{fcurrent} in short) is defined as a continuous linear form on $\Omega_{0}^{p}(E,M)$ for the uniform norm. The space of functional $p$-current will be therefore denoted $\Omega_{0}^{p}(E,M)'$. 
\end{de}
Just as one can establish a correspondence between shapes and currents, to any functional shape we now associate a fcurrent.
\begin{prop}
 Let $(X,f)$ be a functional shape, with $X$ an oriented sub-manifold of dimension $d$ and of finite volume and $f$ a measurable function from $X$ to $M$. For all $\omega \in \Omega_{0}^{d}(E,M)$, $x \mapsto \omega_{(x,f(x))}$ can be integrated along $X$. We set :
\begin{equation}
\label{funshape_fcurrent}
 C_{(X,f)}(\omega):=\int_{X} \omega_{(x,f(x))}\,. 
\end{equation} 
Then $C_{(X,f)}\in \Omega_{0}^{d}(E,M)'$ and therefore $ (X,f) \mapsto C_{(X,f)}$ associates, to any functional shape, a functional current.
\end{prop}
To be more explicit, recall that the integral in \eqref{funshape_fcurrent} is simply defined through local parametrization with a given partition of the unit of sub-manifold $X$. If $F:U \rightarrow E$ is a parametrization of $X$ with $U$ an open subset of $\mathbb{R}^{d}$, then
\begin{equation*}
 C_{(X,f)}(\omega)=\int_{(x_{1},..,x_{d}) \in U} \omega_{(F(x_{1},..,x_{d}),f\circ F(x_{1},..,x_{d}))} \left ( \frac{\partial F}{\partial x_{1}} \wedge ... \wedge \frac{\partial F}{\partial x_{d}} \right ) dx_{1}...dx_{d}\,.
\end{equation*}
Note also, although we did not state it explicitly, that the previous proposition could include sub-manifolds with boundary in the exact same way since the boundary is of zero Hausdorff measure on the sub-manifold. Of course, like for regular currents, the previous correspondence between functional shapes and functional currents is not surjective. For instance, a sum of functional currents of the form $C_{(X,f)}$ do not generally derive from a functional shape. In the functional current framework, Dirac masses are naturally generalized by elementary functional currents or Dirac fcurrents $\delta_{(x,m)}^{\xi}$ for $x \in X, \ m \in M$ and $\xi \in \bigwedge^{p}E$ such that $\delta_{(x,m)}^{\xi}(\omega)=\omega_{(x,m)}(\xi)$. In the same way as explained in the previous part, one can give a discretized version of functional currents associated to $(X,f)$ when a mesh is defined on $X$. $C_{(X,f)}$ is then approximated into a sum of punctual currents :
\begin{equation}
\label{discr_courantsf}
 C_{(X,f)} \approx \sum_{k=1..N} \delta_{(x_{k},m_{k})}^{\xi_{k}}
\end{equation}
In the particular case of a triangulated surface, the discretized version of the fcurrent can be simply obtained as explained for classical currents by adding the interpolated value of signal at each center point of triangles. From the previous equation, we can observe that functional currents have a very simple interpretation. It consists in attaching values of the signal $f$ to the usual representation of $X$ as a $d$-current. At this stage, we could also point out an alternative way to define fcurrents by considering them as tensor products of $d$-currents in $E$ and 0-current (i.e. measure) in $M$, following for instance \cite{DeRham}. 
\subsection{Diffeomorphic transport of fcurrents}
\label{subsec:diff_transp}
What about diffeomorphic transport of functional shapes and currents ? This question cannot be addressed as simply as for the classical current setting if we want to remain completely general. The reason is that, depending on the nature of the signal defined on the manifold, there is not a unique way a deformation can act on a functional shape. In the most simple case where the signal values are not directly correlated to geometry (for instance an activation map on a cortical surface), the natural way to deform a functional shape $(X,f)$ by a diffeomorphism $\phi$ is to transport the geometry of the shape with the values of the signal unchanged. Therefore, the image of $(X,f)$ would be $(\phi(X),f \circ \phi^{-1})$. But imagine now that $f$ is a tangent vector field on $X$. A diffeomorphism $\phi$, by transporting the geometrical support also has to act on the signal through its differential in order to have a tangent vector field on the image shape. In this case, the image of $(X,f)$ is $(\phi(X),g)$ where, for all $y \in \phi(X), \ g(y)= d_{\phi^{-1}(y)} \phi (f\circ \phi^{-1}(y))$. In other cases, for instance a tensor field defined on a manifold, the expression of the transport would differ again. In all cases though, what we have is a left group action of diffeomorphisms of $E$ on the set of considered functional shapes. \\
Thus, to remain general, suppose that a certain class of functional shapes together with such a group action are fixed, we will note $\phi.(X,f)$ the action of $\phi \in \Diff(E)$ on a functional shape $(X,f)$. Then, 
\begin{de}
We call a deformation model on the space of functional currents an action of the group of diffeomorphisms of $E$ on $\Omega_{0}^{d}(E,M)'$ which is such that for any functional shape $(X,f)$ and any diffeomorphism $\phi$, if $\phi_{\ast}$ stands for the action on fcurrents, the following property holds :
\begin{equation}
\label{transport_fcurrent}
 [\phi_{\ast}C_{(X,f)}](\omega)=C_{\phi.(X,f)}(\omega)
\end{equation}
for all $\omega \in \Omega_{0}^{d}(E,M)$. 
\end{de}
Note the difference with \eqref{push_courants} : the action of a diffeomorphism on usual currents is always the simple push forward operation which is automatically compatible with the transport of a shape. Here, it is necessary to adapt the definition of the action on fcurrents to be compatible with a given action on functional shapes by satisfying \eqref{transport_fcurrent}. 

In practical applications, this is usually not a difficulty. In the first case mentioned above, the action of $\phi \in \Diff(E)$ on a functional current $C$ can be derived in a very similar way to the case of usual currents : 
\begin{equation}
\label{eq:trivial_action}
\left\{
  \begin{array}[h]{l}
     \phi_{\ast}C(\omega) \doteq C(\phi^{\ast}\omega), \ \forall \omega \in \Omega_{0}^{d}(E,M)\\
\\
\text{where for all } \forall x \in E, \ m \in M, \ \xi=\xi_{1}\wedge...\wedge \xi_{p} \in \bigwedge^{p}E\\
\\
 (\phi^{\ast}\omega)_{(x,m)}(\xi) \doteq \omega_{(\phi(x),m)}(d_{x}\phi(\xi_{1})\wedge ... \wedge d_{x}\phi(\xi_{d}))\,.
  \end{array}\right.
\end{equation}
It can be easily checked from the previous equations that for all functional shape $(X,f)$, we have $\phi_{\ast}C_{(X,f)}=C_{(\phi(X),f\circ \phi^{-1})}$ as we expected under this model. Since we do not want to focus this paper specifically on deformation, the examples of matching that we will give in the last section are under the hypothesis of this model of transport, which is the simplest and will lead to a convenient generalization of matching algorithms on functional currents. We could go a step further and introduce also a contrast change $\psi\mapsto \psi\circ f$ for $\psi\in \text{Diff}(M)$ so that we end up with a new action of $\text{Diff}(E)\times\text{Diff}(M)$ on $\Omega_0^d(E,M)$ defined by 
\begin{equation}
((\phi,\psi)^\ast\omega)_{(x,m)}(\xi)\doteq \omega_{(\phi(x),\psi(m))}(d_{x}\phi(\xi_{1})\wedge ... \wedge d_{x}\phi(\xi_{d}))\label{eq:phi,psi}
\end{equation}
and the corresponding action on fcurrent $  (\phi,\psi)_{\ast}C(\omega) \doteq C((\phi,\psi)^{\ast}\omega)$ given by duality for which we easily check that
\begin{equation}
(\phi,\psi)_{\ast} (\delta_{x,m}^\xi)=\delta_{\phi(x),\psi(m)}^{d_{x}\phi(\xi_{1})\wedge ... \wedge d_{x}\phi(\xi_{d})}\label{eq:phi,psi,dual}\,.
\end{equation}
Note that it is not significantly more difficult to express and implement the deformation model on functional currents that corresponds to other types of action, as for instance in the case of tangent vector signal we mentioned earlier.

\section{A Hilbert space structure on functional currents} 
\label{sec:hilbert}
In this section, we address the fundamental question of \emph{comparing} functional currents through an appropriate metric. For this purpose, we adapt the ideas of RKHS presented briefly for currents in the first part of the paper. This approach allows to view functional currents as elements of a Hilbert space of functions, which opens the way to various processing algorithms on functional shapes as will be illustrated in the next section.
\subsection{Kernels on fcurrent spaces}
As we have seen for currents, the theory of RKHS defines an inner product between currents through a certain kernel function satisfying some regularity and boundary conditions. Following the idea that functional $p$-currents can be considered as well as tensor product of $p$-currents on E and $0$-currents on M, we can generically define a kernel on $E \times M$.
\begin{prop}
\label{propfkernel}
 Let $K_{g}: E \times E \rightarrow \mathcal{L}(\bigwedge^{p}E)$ be a positive kernel on the geometrical space $E$ and $K_{f}: M \times M \rightarrow \mathbb{R}$ a positive kernel on the signal space $M$. We assume that both kernels are continuous, bounded and vanishing at infinity. Then $K_{g} \otimes K_{f}$ defines a positive kernel from $E \times M$ on $\bigwedge^{p}E$ whose corresponding reproducing Hilbert space $W$ is continuously embedded into $\Omega_{0}^{p}(E,M)$. Consequently, every functional p-current belongs to $W'$.
\end{prop}
\begin{proof} This relies essentially on classical properties of kernels. From the conditions on both kernels, we know that to $K_{g}$ and $K_{f}$ correspond two RKHS $W_{g}$ and $W_{f}$ that are respectively embedded into $\Omega_{0}^{p}(E)$ and $\Omega_{0}^{0}(M)$ (cf \cite{Glaunes}). It is a classical result in RKHS theory that $K:= K_{g} \otimes K_{f}$ defines a positive kernel. Moreover, since $K_{f}$ is real-valued, we have the following explicit expression of $K$:
\begin{equation*}
 K \left( (x_{1},m_{1}),(x_{2},m_{2}) \right ) = K_{f}(m_{1},m_{2}).K_{g}(x_{1},x_{2})\,.
\end{equation*}
To the kernel $K$ corresponds a unique RKHS $W$ that is the completion of the vector space spanned by all the functions $\{K_{f}(.,m).K_{g}(.,x)\xi\}$ for $x\in E, \ m \in M, \ \xi \in \bigwedge^{p}E$. Since functions $K_{f}(.,m)$ and $K_{g}(.,x)$ are both continuous and vanishing at infinity from what we have said, it is also the case for $K_{f}(.,m).K_{g}(.,x)\xi$ so that $W$ is indeed embedded into $\Omega_{0}^{p}(E,M)$. There only remains to prove that we have a \emph{continuous} embedding, which reduces to dominate the uniform norm by $\|.\|_{W}$.  

Let $\omega \in W$. For all $(x,m) \in E \times M$ and $\xi \in \bigwedge^{p}E$ such that $|\xi|=1$, we have
\begin{equation}
\label{intermediate1}
 |\omega_{(x,m)}(\xi)| = |\delta_{(x,m)}^{\xi}(\omega)| \,.
\end{equation}
Since $W$ is a RKHS, all $\delta_{(x,m)}^{\xi}$ are continuous linear forms on $W$. In addition, Riesz representation theorem provides an isometry $K_{W} : W' \rightarrow W$. Then :
\begin{eqnarray}
\label{ps_courantsf1}
 \langle \delta_{(x_{1},m_{1})}^{\xi_{1}} , \delta_{(x_{2},m_{2})}^{\xi_{2}} \rangle_{W'} &=&  \langle K_{W}(\delta_{(x_{1},m_{1})}^{\xi_{1}}) , K_{W}(\delta_{(x_{2},m_{2})}^{\xi_{2}}) \rangle_{W} \nonumber \\
&=& \langle K_{f}(.,m_{1})K_{g}(.,x_{1})\xi_{1} , K_{f}(.,m_{2})K_{g}(.,x_{2})\xi_{1}  \rangle_{W} \nonumber \\
&=& K_{f}(m_{1},m_{2}). \xi_{2}^{T} K_{g}(x_{1},x_{2}) \xi_{1}
\end{eqnarray} 
Now, back to equation (\ref{intermediate1}), we have :
\begin{eqnarray*}
 |\omega_{(x,m)}(\xi)| &\leq& \|\delta_{x,m}^{\xi}\|_{W'} \ \|\omega\|_{W} \\
&\leq& \sqrt{K_{f}(m,m).\xi^{T}K_{g}(x,x)\xi} \ \|\omega\|_{W}
\end{eqnarray*}
Since we assume that $m\mapsto K_{f}(m,m)$ and $x\mapsto K_{g}(x,x)$ are bounded we deduce that $\sqrt{K_{f}(m,m).\xi^{T}K_{g}(x,x)\xi}$ is bounded with respect to $x$, $m$ and $\xi$ with $|\xi|=1$. Hence, by taking the supremum in the previous equation, we finally get
\begin{eqnarray*}
 \|\omega\|_{\infty} \leq C \|\omega\|_{W}
\end{eqnarray*}
which precisely means that the embedding is continuous. By duality, we get that every functional current is an element of $W'$. Note that the dual application is not necessarily injective unless $W$ is dense in $\Omega_{0}^{p}(E,M)$, which is the case in particular if both $W_{g}$ and $W_{f}$ are respectively dense in $\Omega_{0}^{p}(E)$ and $\Omega_{0}^{0}(M)$. \end{proof}
In other words, a quite natural (but not unique) way to build kernels for functional currents is to make the tensor product of kernels defined separately in the geometrical domain ($p$-currents in $E$) and in the signal domain ($0$-currents in $M$). As we see, everything eventually relies on the specification of kernels on $E$ and $M$. 

Kernels on vector spaces have been widely studied in the past and obviously do not arise any additional difficulty in our approach compared to usual current settings. Among others, classical examples of kernels on a vector space $E$ taking values in another vector space $H$ are provided by radial scalar kernels defined for $x,y \in E$ by $K(x,y)=k(|x-y|).\text{Id}_{H}$ where $k$ is a function defined on $\mathbb{R_{+}}$ and vanishing at infinity. This family of kernels is the only one that induces a RKHS norm invariant through affine isometries. The most popular is the Gaussian kernel defined by $K(x,y)= \exp \left ( -\frac{|x-y|^{2}}{\sigma^{2}} \right )\text{Id}_{H}$, $\sigma$ being a scale parameter that can be interpreted somehow as a range of interactions between points. 

The definition of a kernel on a general manifold $M$ is often a more involved issue as we already mentioned in subsection \ref{subsec:func}. Generally, the procedure is reversed : the kernel is defined through a compact operator on differential forms of $M$, which can be diagonalized and hopefully provide a closed expression of the kernel on $M$ (cf \cite{Zeidler}). The case of the two-dimensional sphere for instance is thoroughly treated in \cite{Glaunes}. However, it's important to note that, in our setting of functional currents, this issue is drastically simplified because we only need to define a \textbf{real-valued} kernel on $M$. This is  contrasting with the idea of product space currents of subsection \ref{subsec:func}, which requires the definition of kernels living in the exterior product of the fiber bundle of $M$. For instance, if $M$ is a sub-manifold of a certain vector space, obtaining real-valued kernels on $M$ becomes straightforward by restriction to $M$ of kernels defined on the ambient vector space.


\subsection{Convergence and control results on the RKHS norm}
\label{subsec:control}
We are now going to explore a little more some properties of the RKHS norm on fcurrents and show the theoretical benefits of our approach with respect to the original problem raised by this article. 

Suppose, under the same hypotheses as the previous section, that two kernels $K_{g}$ and $K_{f}$ are given respectively on space $E$ and manifold $M$, providing two RKHS $W_{g}$ and $W_{f}$. By a simple triangular inequality, we get for any $x_1$, $x_2 \in E$, any $\xi_1$, $\xi_2 \in \bigwedge^{p} E$ and any $m_1$, $m_2 \in M$ 
\begin{equation}
\label{inegalite_distance}
  \|\delta_{(x_{2},m_{2})}^{\xi_{2}} - \delta_{(x_{1},m_{1})}^{\xi_{1}}\|_{W'} \leq \| \delta_{m_{1}} \|_{W_{f}'} \| \delta_{x_{2}}^{\xi_{2}} - \delta_{x_{1}}^{\xi_{1}} \|_{W_{g}'} + \| \delta_{x_{2}}^{\xi_{2}} \|_{W_{g}'} \| \delta_{m_{2}} - \delta_{m_{1}} \|_{W_{f}'}\,.
\end{equation}
Since both kernels $K_{f}$ and $K_{g}$ are assumed to be bounded as in Proposition \ref{propfkernel}, $\| \delta_{m_{1}} \|_{W_{f}'}$ and $\| \delta_{x_{2}}^{\xi_{2}} \|_{W_{g}'}$ are uniformly bounded so that eventually
\begin{equation*}
\|\delta_{(x_{2},m_{2})}^{\xi_{2}} - \delta_{(x_{1},m_{1})}^{\xi_{1}}\|_{W'} \leq \text{Cst}\,(\| \delta_{x_{2}}^{\xi_{2}} - \delta_{x_{1}}^{\xi_{1}} \|_{W_{g}'} + \| \delta_{m_{2}} - \delta_{m_{1}} \|_{W_{f}'})\,.
\end{equation*}
Therefore, the RKHS distance between punctual fcurrents is dominated both with respect to the variation of their geometrical parts and of their functional values. This is the general idea we will formulate in a more precise way with the two following propositions. We denote by $d_{M}$ the geodesic distance induced on $M$ by its Riemannian structure. The next proposition examines the case where the geometrical support is a fixed sub-manifold $X$ and shows that the variation of the $W'$-norm is then dominated by the $L^{1}$ norm on $X$.  
\begin{prop}
\label{propcontrolnorm}
Let $X$ be a d-dimensional sub-manifold of $E$ of finite volume and $f_{1}:X \rightarrow M$ and $f_{2}: X \rightarrow M$ two measurable functions defined on the sub-manifold $X$ taking value in $M$. We assume that $W_{f}$ is continuously embedded into $C_{0}^{1}(M,\mathbb{R})$. Then, there exists a constant $\beta$ such that :
\begin{equation*}
 \|C_{(X,f_{1})} - C_{(X,f_{2})}\|_{W'}\leq \beta \int_{X} d_{M}(f_{1}(x),f_{2}(x)) d\sigma(x)\,
\end{equation*}
where $\sigma$ is the uniform measure on $X$.
\end{prop} 
\begin{proof} We recall the definition $C_{(X,f)}=\int_{X} \omega_{(x,f(x))}$. We will first restrict the proof to the case where $X$ admits a parametrization given by a function $G:U\rightarrow E$ where U is an open subset of $\mathbb{R}^{d}$. The general result follows by the use of an appropriate partition of the unit on $X$. Denoting $\xi(u)\doteq \frac{\partial G}{\partial u_{1}}(u)\wedge..\wedge\frac{\partial G}{\partial u_{d}}(u)$ for $u=(u_{1},..,u_{d})\in U$, we get
\begin{equation*}
 C_{(X,f)}(\omega)= \int_{u\in U} \omega_{(G(u),f\circ G(u))}(\xi(u)) du
\end{equation*}
Now, for $g_{1}\doteq f_{1}\circ G$ and $g_{2}\doteq f_{2}\circ G$, we have by triangular inequality on $\|.\|_{W'}$~:
\begin{equation}
\label{propcontrolnorm_int1}
\|C_{(X,f_{1})} - C_{(X,f_{2})}\|_{W'} \leq \int_{U} \|\delta_{(G(u),g_{1}(u))}^{\xi(u)}-\delta_{(G(u),g_{2}(u))}^{\xi(u)}\|_{W'} du\,.
\end{equation}
From \eqref{inegalite_distance}, $\|\delta_{(G(u),g_{1}(u))}^{\xi(u)}-\delta_{(G(u),g_{2}(u))}^{\xi(u)}\|_{W'} \leq \| \delta_{G(u)}^{\xi(u)} \|_{W_{g}'} \|\delta_{g_{2}(u)} - \delta_{g_{1}(u)} \|_{W_{f}'}$. Now, for any $m_{1}$, $m_2\in M$ and $h\in W_f$ we have 
\begin{eqnarray*}
 |(\delta_{m_{1}}-\delta_{m_{2}})(h)| &=& |h(m_{1})-h(m_{2})| \\
&\leq& \|Dh\|_{\infty} d_{M}(m_{1},m_{2}) \\
&\leq& \text{Cst}\,\|h\|_{W_{f}} d_{M}(m_{1},m_{2})
\end{eqnarray*}
the last inequality resulting from the continuous embedding $W_{f}\hookrightarrow C_{0}^{1}(M,\mathbb{R})$. Therefore we get
\begin{equation*}
 \|\delta_{g_{2}(u)} - \delta_{g_{1}(u)} \|_{W_{f}'} \leq \text{Cst}\,d_{M}(g_{1}(u),g_{2}(u))\,.
\end{equation*}
Moreover, since we assume that the kernel $K_{g}$ is bounded, we also have $\| \delta_{G(u)}^{\xi(u)} \|_{W_{g}'} \leq \text{Cst}\,|\xi(u)|$. Back to equation \eqref{propcontrolnorm_int1}, we get from the previous derivations the existence of a constant $\beta>0$ such that~:
\begin{equation*}
 \|C_{(X,f_{1})} - C_{(X,f_{2})}\|_{W'} \leq \beta \int_{U} d_{M}(g_{1}(u),g_{2}(u)) |\xi(u)| du
\end{equation*}
which precisely proves the stated result. \end{proof}
A straightforward consequence of Proposition \ref{propcontrolnorm} and dominated convergence theorem is that if $f_{n}$ is a sequence of function on $X$ that converges pointwisely to a function $f$, then $C_{(X,f_{n})} \rightarrow C_{(X,f)}$. In other words, \textbf{pointwise convergence of signal implies convergence in terms of fcurrents}. \\

Following the same kind of reasoning we eventually give a local bound of the RKHS distance between a functional shape and the same shape deformed through small diffeomorphisms both in geometry and signal. As it is now classical, we consider deformations modelled as flows between 0 and 1 of differential equations given through time varying vector fields. In appendix A, we remind the basic definitions about this modelling and a few useful results for the following. Let $u(t,x)$ (resp. $v(t,m)$) be a smooth time dependent vector fields on the geometrical space $E$ (resp. on the signal space $M$) and let $\phi$ (resp. $\psi$) the solution at time $1$ of the flow of the ODE $y'=u(t,y)$ (resp. $y'=v(t,y)$). On these spaces of vector fields, we define the norms :
\begin{equation*}
 \| u \|_{\chi^{1}} = \int_{0}^{1} |u(t,.)|_{1,\infty} dt, \ \| v \|_{\chi^{0}} = \int_{0}^{1} |v(t,.)|_{0,\infty} dt
\end{equation*}
where $|u(t,.)|_{1,\infty}=\sup_{x}|u(t,x)|+\sum_{i}\sup_{x}|\frac{\partial u}{\partial x_i}(t,x)|$ and $\|v(t,.)\|_{0,\infty}=\sup_{m}|v(t,m)|$.
\begin{prop}
\label{propcontrolnormdefor}
Let $X$ be a sub-manifold of $E$ of finite volume and $f: X\rightarrow M$ a measurable function. Assume that $W_{g}$ and $W_{f}$ are continuously embedded respectively into $C_{0}^{1}(E,\bigwedge^{d}E^{*})$ and $C_{0}^{1}(M,\mathbb{R})$. There exists a universal constant $\gamma > 0$ such that, if $\| u \|_{\chi^{1}}$ and $\| v \|_{\chi^{0}}$ are sufficiently small (which means that deformations are 'close' to identity), then : 
\begin{equation*}
 \|C_{(X,f)} - C_{(\phi(X),\psi \circ f \circ \phi^{-1})}\|_{W'} \leq \gamma  \text{Vol}(X) \left ( \|u\|_{\chi^{1}} + \|v\|_{\chi^{0}} \right )
\end{equation*}
\end{prop}
\begin{proof} The full proof of proposition \ref{propcontrolnormdefor} relies mostly on a few controls which are summed up in appendix A. Given again a local parametrization of $X$, $G: U \rightarrow X$, then, similarly to the previous proposition and using same notations, we have :
\begin{equation}
\label{propcontrolnormdefor_inter1}
 \|C_{(X,f)} - C_{(\phi(X),\psi \circ f \circ \phi^{-1})}\|_{W'} \leq \int_{U} \|\delta_{(G(u),f\circ G(u))}^{\xi(u)}-\delta_{(\phi \circ G(u), \psi \circ f \circ G(u))}^{\tilde{\xi}(u)}\|_{W'} du 
\end{equation}
where for the volume element $\xi(u)=\xi_{1}(u)\wedge...\wedge \xi_{d}(u)$, $\tilde{\xi}(u)$ is the transported volume element by $\phi$ equal to $\tilde{\xi}(u)=d\phi_{x}(\xi_{1}(u)) \wedge...\wedge d\phi_{x}(\xi_{d}(u))$. From \eqref{inegalite_distance} we get
\begin{equation*}
  \begin{split}
    \|\delta_{(x,f(x))}^{\xi(x)}-\delta_{(\phi(x), \psi \circ f(x))}^{\tilde{\xi}(x)}\|_{W'} & \leq \|\delta_{\psi \circ f(x)}\|_{W_{f}'} \|\delta_{\phi(x)}^{\tilde{\xi}(x)} - \delta_{x}^{\xi(x)}\|_{W_{g}'}\\
& + \|\delta_{x}^{\xi(x)}\|_{W_{g}'} \|\delta_{\psi \circ f(x)} - \delta_{f(x)}\|_{W_{f}'}\,.
  \end{split}
\end{equation*}
and using $\|\delta_{x}^{\xi(x)}\|_{W_{g}'} \leq \text{Cst}\,|\xi(x)|$ and $\|\delta_{\psi \circ f(x)} - \delta_{f(x)}\|_{W_{f}'} \leq \text{Cst}\,d_{M}(\psi \circ f(x), f(x))$ we get
\begin{equation}
\label{propcontrolnormdefor_inter2}
\begin{split}
  \|\delta_{x}^{\xi(x)}\|_{W_{g}'} \|\delta_{\psi \circ f(x)} -
  \delta_{f(x)}\|_{W_{f}'} & \leq \text{Cst}\,|\xi(x)| d_M(\psi\circ
  f(x),f(x))\\
&\leq \text{Cst}\,|\xi(x)|\|v\|_{\chi^{0}}
\end{split}
\end{equation}
In a similar way, we know that $\|\delta_{\psi \circ f(x)}\|_{W_{f}'} \leq \text{Cst}$. Moreover :
\begin{eqnarray*}
\|\delta_{x}^{\xi(x)} - \delta_{\phi(x)}^{\tilde{\xi}(x)}\|_{W_{g}'} &\leq& \text{Cst}\,(|\xi(x)|\|Id - \phi \|_{\infty} + |\tilde{\xi}(x) - \xi(x)|) \\
&\leq& \text{Cst}\,|\xi(x)| \|u\|_{\chi^{1}}
\end{eqnarray*}
the last inequality being obtained thanks to theorem \ref{cor_diffeo} and corollary \ref{cor_jac_dvect} of appendix A with $s=0$ and $t=1$. This leads to :
\begin{equation}
\label{propcontrolnormdefor_inter3}
\|\delta_{\psi \circ f(x)}\|_{W_{f}'} \|\delta_{x}^{\xi(x)} - \delta_{\phi(x)}^{\tilde{\xi}(x)}\|_{W_{g}'} \leq \text{Cst}\,|\xi(x)| \|u\|_{1,\infty}\,.
\end{equation}
Plugging \eqref{propcontrolnormdefor_inter2} and \eqref{propcontrolnormdefor_inter3} in \eqref{propcontrolnormdefor_inter1}, we finally get :
\begin{equation*}
\|C_{(X,f)} - C_{(\phi(X),\psi \circ f \circ \phi^{-1})}\|_{W'} \leq \text{Cst}\,(\|u\|_{\chi^{1}}+\|v\|_{\chi^{0}}) \int_{U} |\xi(u)| du
\end{equation*}
which concludes the proof since $\int_{U} |\xi(u)| du = \text{Vol}(X)$. 
\end{proof}
This property shows that the RKHS norm is continuous with respect to deformations of the functional shape (both in its geometry and its signal). More specifically, it is not hard to see that $C_{(\phi(X),\psi\circ f\circ \phi^{-1})}=(\phi,\psi)_\ast C_{(X,f)}$ for the action given by (\ref{eq:phi,psi}) and (\ref{eq:phi,psi,dual}) and to extend the proof of the previous proposition to a more general situation of a fcurrent $C\in W'$ having finite ``mass norm'' $M(C)$ where $M(C)\doteq \sup_{\omega\in W,\|\omega\|_\infty\leq 1} C(\omega)$ is the proper extension of the previous finite volume condition. Then we get
\begin{equation}
  \label{eq:cont_action}
  \|(\phi,\psi)_\ast C - C\|_{W'}\leq \gamma M(C)\left ( \|u\|_{\chi^{1}} + \|v\|_{\chi^{0}} \right )\,.
\end{equation}
where $\gamma$ is a universal constant.

\begin{figure}
\center
   \includegraphics[width=8cm]{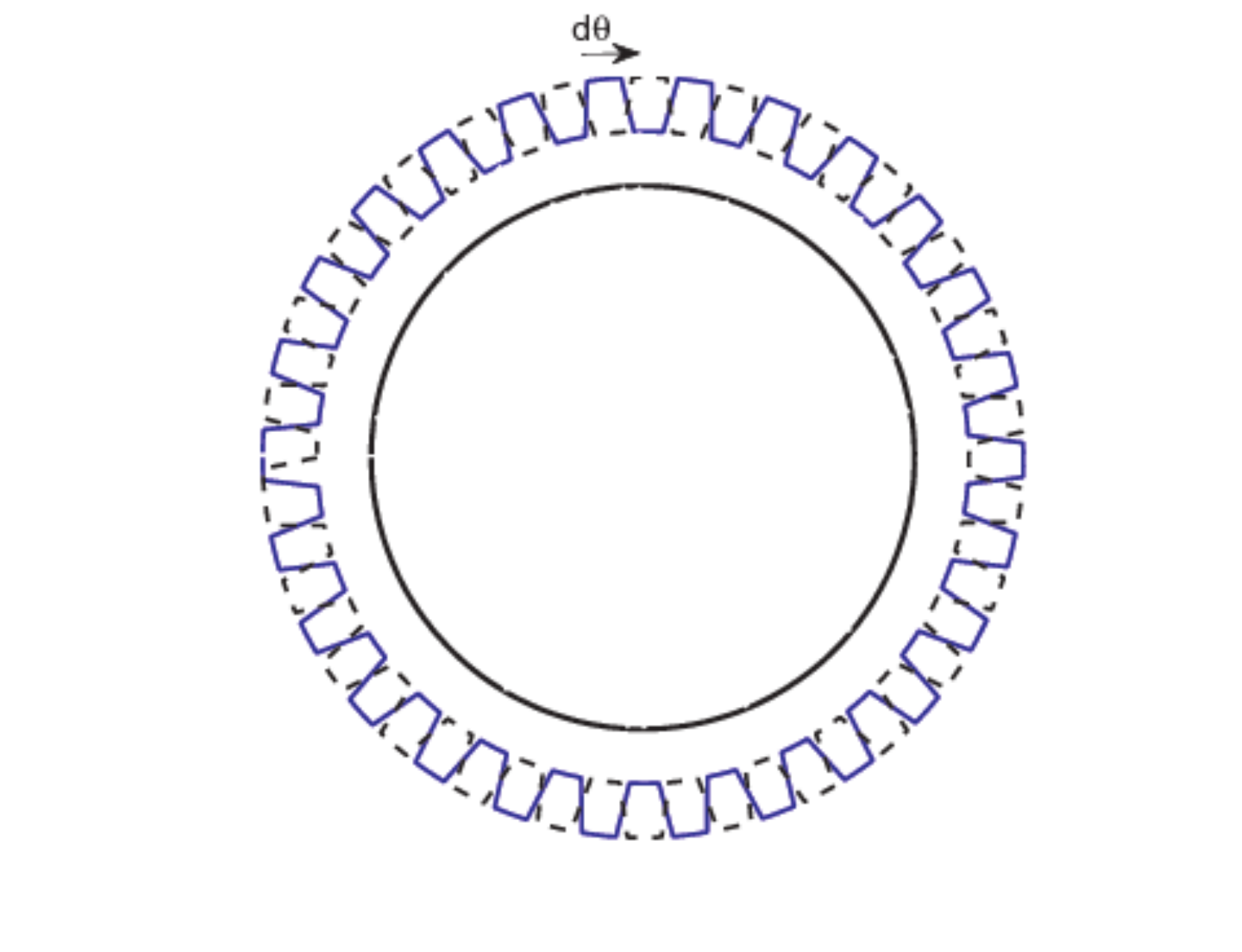} 
\caption{Comparison of the fcurrent's norm and the $L^{p}$ norm on a fixed geometrical support : example of crenellations on the unit circle.}
\label{comp_Lp_RKHS}
\end{figure}

This result also provides an answer to wether there is a resversed domination in Proposition \ref{propcontrolnorm} for two functional shapes that have the same geometrical support. Indeed, consider a particular case where $\psi=\text{Id}$ and $\phi$ is a small deformation that leaves $X$ globally invariant ($\phi(X)=X$). We wish to compare the initial functional shape $(X,f)$ with the deformed one $(\phi(X),f\circ\phi^{-1})=(X,f \circ \phi^{-1})$. By proposition \ref{propcontrolnormdefor}, we know that, for any function $f$, the fcurrent's distance remains small if the deformation $\phi$ is small. It is no longer true if we compute instead $\int_{X} |f-f \circ \phi^{-1}|^{p}$, the $L^{p}$ distance on $X$ ($0< p \leq \infty$). This is easily seen if we choose for $X$ the unit circle $\mathbb{S}^{1}$ and consider crenellated signals as in figure \ref{comp_Lp_RKHS}. Introducing the operator $\tau_{d\theta}$ that acts on functional shapes by rotation of an angle $d\theta$, we see indeed that : 
\begin{equation*}
\sup_{f \in L^{^{p}} (\mathbb{S}^{1}),\|f\|_{L^{p}} \leq 1} \int_{\mathbb{S}^{1}} |f-f \circ \tau_{d\theta}^{-1}|^{p} = 1
\end{equation*}
whereas, according to Proposition \ref{propcontrolnormdefor} 
\begin{equation*}
\sup_{f \in L^{^{p}} (\mathbb{S}^{1}),\|f\|_{L^{p}} \leq 1} \|C_{(X,f)} - C_{(X, f \circ \tau_{d\theta}^{-1})}\|_{W'} = O(d\theta)\,.
\end{equation*}
In conclusion, this gives an answer to the previous question : $W'$ norm and $L^{p}$ norm on a fixed geometrical support are \textbf{not} equivalent in general. Again, such a fact speaks in favor of the use of RKHS norms on fcurrents : somehow, the approach we have presented allows a coherent collaboration between signal and geometry to define a proper attachment term for functional shapes that shall be used in section \ref{sec:processing}.

\section{Processing functional shapes with fcurrents : Two examples}
\label{sec:processing}
 We would like to illustrate now how the concept of functional currents introduced before offers a genuine solution to the simultaneous processing of the geometric and signal information of any functional shape. We have explained how functional currents can be equipped with a Hilbertian norm mixing geometrical and functional content of functional shapes and how this norm has nice properties with respect to geometrical and functional perturbations. It is more or less clear that the embedding in this convenient Hilbert setting is opening to way to various processing algorithms that will be developed in a near future. Since the purpose of this paper is to stay focused on the theoretical exposition of the fcurrents, you will not try here to develop a full range of applications but we will present briefly two illustrative applications in order to shed a light on the potential of the developed framework. The first application illustrates the full potential of the Hilbertian structure with the design of redundancy reduction or compression algorithms for functional shapes representations through matching pursuit schemes on functional currents. The second one, closer to the core engine of computational anatomy, is the design of large deformation matching algorithm for the simultaneous geometric and functional registration of functional shapes under diffeomorphic transport.

\subsection{A compression algorithm for functional current representations}
Let us start with the issue of the redundancy of fcurrent representations. If we consider for instance a segment in the 2D space with constant signal, the discretization in punctual fcurrents given by \eqref{discr_courantsf} will provide a representation with a number of elements that corresponds to the initial sampling of the curve. Generally, this representation could be clearly reduced since, for such a simple functional shape, only a few terms should capture \emph{most of} the shape. However, the quality of the approximation needs to be quantified in a meaningful way, especially when the functional part is also involved, through an appropriate norm for which we have a natural candidate given by the Hilbert structure. This issue of redundancy reduction or compression is important for instance when making means of currents because without further treatment, the number of Dirac currents involved in the representation of the mean would increase dramatically. This is even more important when considering higher order statistics for the estimation of noise or texture models around a mean functional shape possibly coupled with a deformation model learned from a set of inexact geodesic matchings, as provided for instance by the matching algorithm provided in subsection \ref{subsec:match}. 
In the following, we only provide a general overview of the algorithm and few numerical results to show the functional current behaviours. The details of numerical optimization that may deserve a more in depth exposition are out of the scope of the present paper.

As we have said, the problem of redundancy reduction or compression is deeply simplified thanks to the Hilbert space structure that has been defined on functional currents in the previous section. Indeed, classical matching-pursuit algorithms in general Hilbert spaces have already been studied by Mallat and Zhang in \cite{Mallat} and later adapted to currents in \cite{Durrlemann2}. We can proceed in a similar way for functional currents. Consider again a discretized fcurrent $C=\sum_{i=1..N} \delta_{(x_{i},m_{i})}^{\xi_{i}} \in W'$. N, the number of momenta, is automatically given by the mesh on the sub manifold (point sampling for curves, triangulation for surfaces,...). This sub manifold might have some very regular regions with low geometrical and functional variations, in which results a very redundant representation by fcurrent due to the fact that many adjacent nodes present the same local geometry and signal. The goal of matching-pursuit is to find a more appropriate and reduced representation of $C$ in elementary functional currents. Given a certain threshold $\varepsilon > 0$, we want to find $\Pi_{n}(C)$ such that $C=\Pi_{n}(C)+R_{n}(C)$ and $\|R_{n}(C)\|_{W'} \leqslant \varepsilon$. $R_{n}(C)$ will be called the residue of the approximation. Somehow, this is linked to the problem of finding the best projection of $C$ on a subspace of $W'$. This problem is however too much time-consuming computationally for usual applications. Instead, matching pursuit is a greedy algorithm that constructs a family of approximating vectors step by step. The result is a suboptimal fcurrent that approximates the functional current C with a residue whose energy is below threshold. The algorithm basically proceeds as follows. We need to specify a 'dictionary' $\mathcal{D}$  of elements in $W'$. In our case, we typically consider the set of all elementary functional currents $\{\delta_{(x,m)}^{\xi}\}$ with $\xi$ a unit vector in $\bigwedge^{d}E$. The first step of matching pursuit algorithm is to find $\delta_{(x'_{1},m_{1}')}^{\xi'_{1}} \in \mathcal{D}$ that is best correlated to C. In other words, we try to maximise, with respect to $x,m,\xi$, the quantity :
\begin{equation}
\label{match_pursuit_innerprod}
 \langle C,\delta_{(x,m)}^{\xi}\rangle_{W'} = \xi^{T} \left ( \sum_{i=1}^{N} K((x,m),(x_{i},m_{i})) \xi_{i} \right )
\end{equation} 
Since $\xi$ is taken among unit vectors, the problem is strictly equivalent to maximize $\|\sum_{i=1..N} K((x,m),(x_{i},m_{i})) \xi_{i} \|=\|\gamma(x,m) \|$ with respect to $(x,m)$ and take $\xi$ as the unit vector of same direction. We get a first approximation of C : 
$$C=\Pi_{1}(C)+R_{1}(C)\,.$$ 
The algorithm then applies the same procedure to the residue $R_{1}(C)$, which provides a second vector $\delta_{(x'_{2},m_{2}')}^{\xi'_{2}} \in \mathcal{D}$, and a residue $R_{2}(C)$. The algorithm is stopped when the RKHS norm of the residue decreases below the given threshold $\varepsilon$. 

In most cases, it appears that the compression is better with the orthogonal version of the previous scheme, in which the family of vectors is orthonormalized at each step, in order to impose the projection and the residue to be orthogonal in $W'$. The classical algorithm is based on a Gram-Schmidt orthonormalization at each step. In our case, it's possible to obtain a similar result in a more optimal way by keeping the values of $(x_{i}',m_{i}')$ found during previous steps and simply modify the vectors $\xi_{i}'$. This is done by imposing the following orthogonality condition. Let's call $(e_{k})$ the canonical basis of the vector space $\bigwedge^{d}E$. If $C=\Pi_{n}(C)+R_{n}(C)$ and $\Pi_{n}(C)=\sum_{i=1..n} \delta_{(x_{i}',m_{i}')}^{\alpha^{n}_{i}}$, we will add the orthogonality constraint :
\begin{equation*}
\delta_{(x_{i}',m_{i}')}^{e_{k}} \bot R_{n}(C) \Longleftrightarrow \langle C,\delta_{(x_{i}',m_{i}')}^{e_{k}}\rangle_{W^{\ast}} = \langle \Pi_{n}(C),\delta_{(x_{i}',m_{i}')}^{e_{k}}\rangle_{W^{\ast}}
\end{equation*} 
for all basis vectors $e_{k}$ and for all $i \in \{1,..,n\}$. It is then straightforward to check that these conditions are strictly equivalent to the following system of linear equations to find the $\alpha^{n}_{i}$ :
 \begin{equation}
\label{syst_lin}
\forall i \in \{1,..,n\}, \ \sum_{j=1}^{n} \left ( K((x_{i}',m_{i}'),(x_{j}',m_{j}'))\alpha^{n}_{j} \right )_{k} = \gamma(x_{i}',m_{i}')_{k}
\end{equation} 
We could show that the norm of the residue $R_{n}(C)$ monotonically decreases to zero as $n\rightarrow \infty$. Hence the algorithm converges and eventually when the residue goes below the given threshold at a certain step $n$, we obtain a compressed representation of $C$ with $n$ orthogonal dirac fcurrents (with generally $n\ll N$, as we shall see on the coming examples). At each step, the time-consuming part of the algorithm is mainly the computation of sums of kernels, which has quadratic complexity with respect to the number of Diracs of the original current but can be speeded up tremendously by making computations on a fixed grid with FFT, as introduced for currents in \cite{Durrlemann}. The same kind of numerical trick can be performed with fcurrents but we will not elaborate on that in this paper.

\begin{figure}[!h]
\leftskip -2.5cm
\begin{tabular}{ccc}
  \includegraphics[width=4.5cm, height=3.6cm]{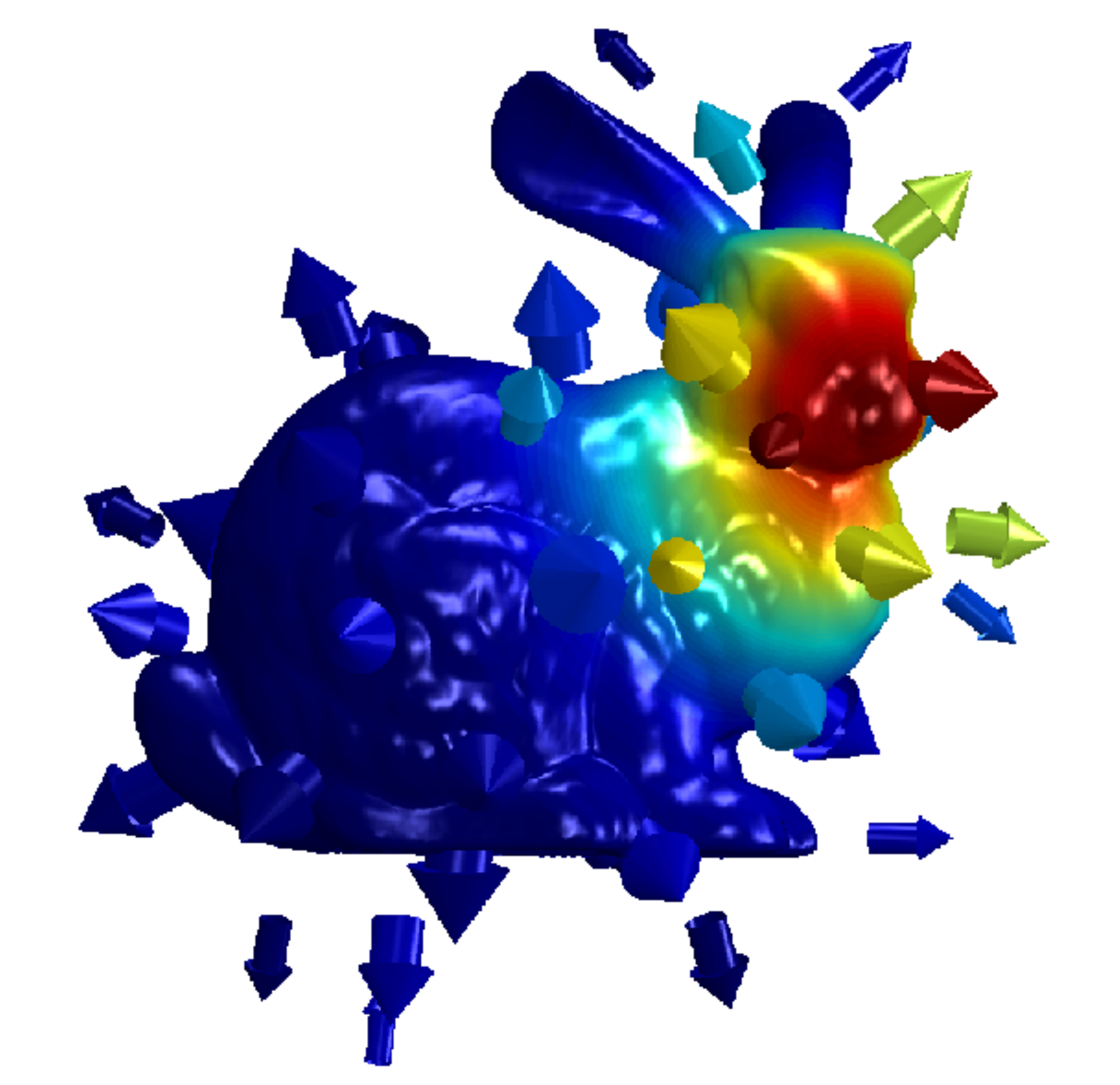} & \includegraphics[width=4.5cm, height=3.6cm]{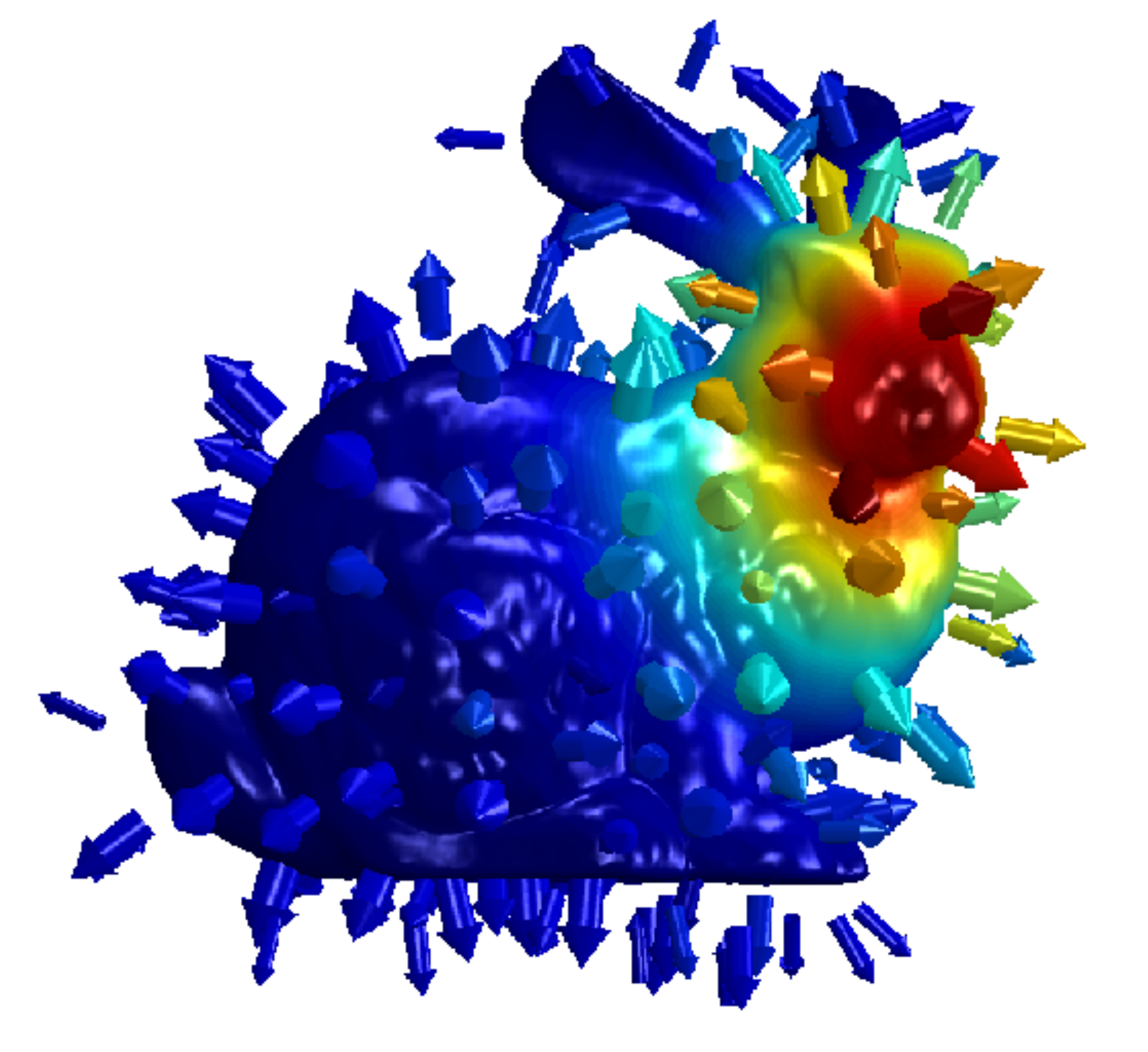} & \includegraphics[width=4.5cm, height=3.6cm]{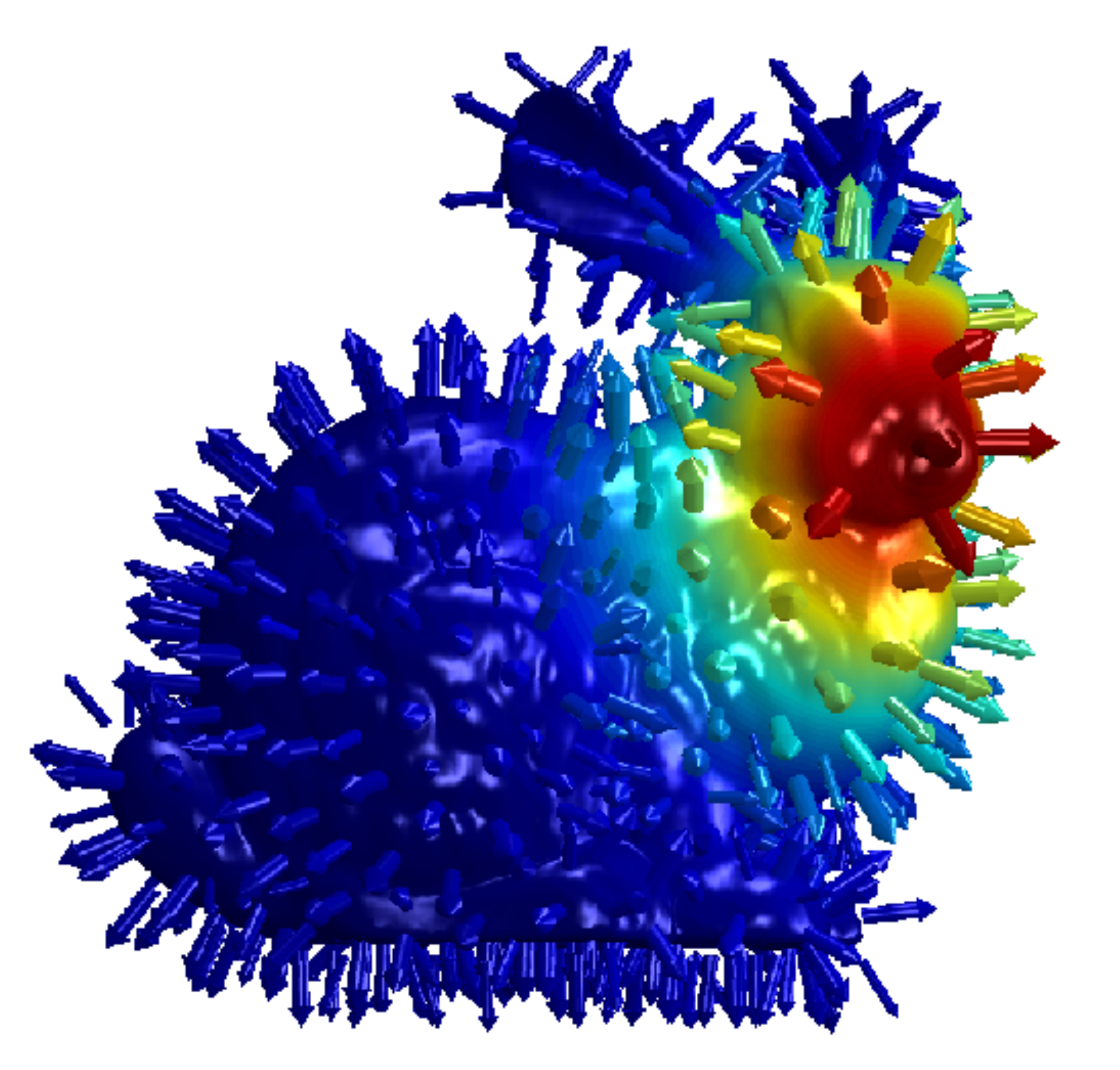} \\
$\lambda_{g}=0.04, \ \lambda_{f}=0.4, \ 47 \ Diracs$ & $\lambda_{g}=0.02, \ \lambda_{f}=0.4, \ 170 \ Diracs$ & $\lambda_{g}=0.01, \ \lambda_{f}=0.2, \ 565 \ Diracs$ \\

  \includegraphics[width=4.5cm, height=3.6cm]{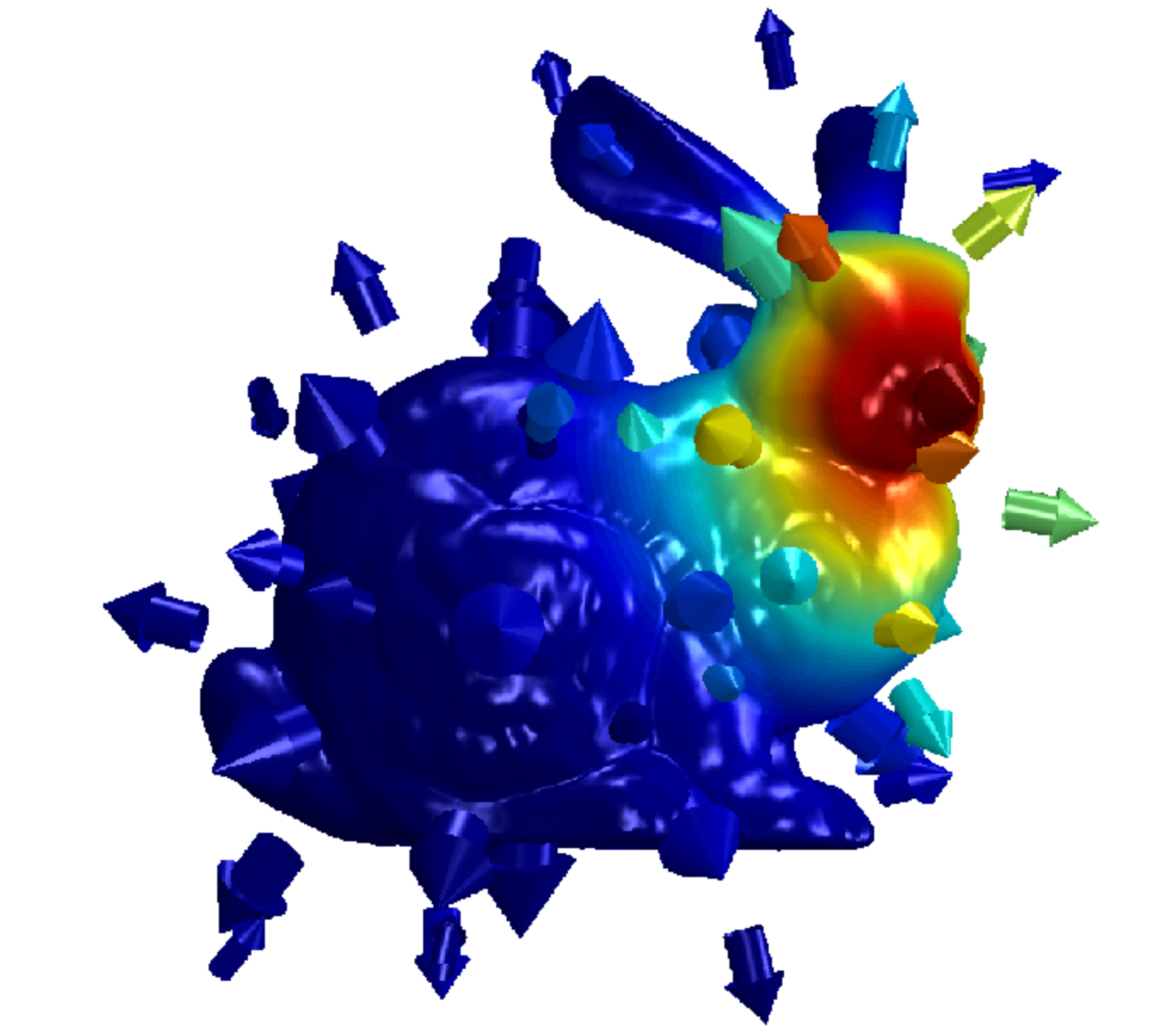} & \includegraphics[width=4.5cm, height=3.6cm]{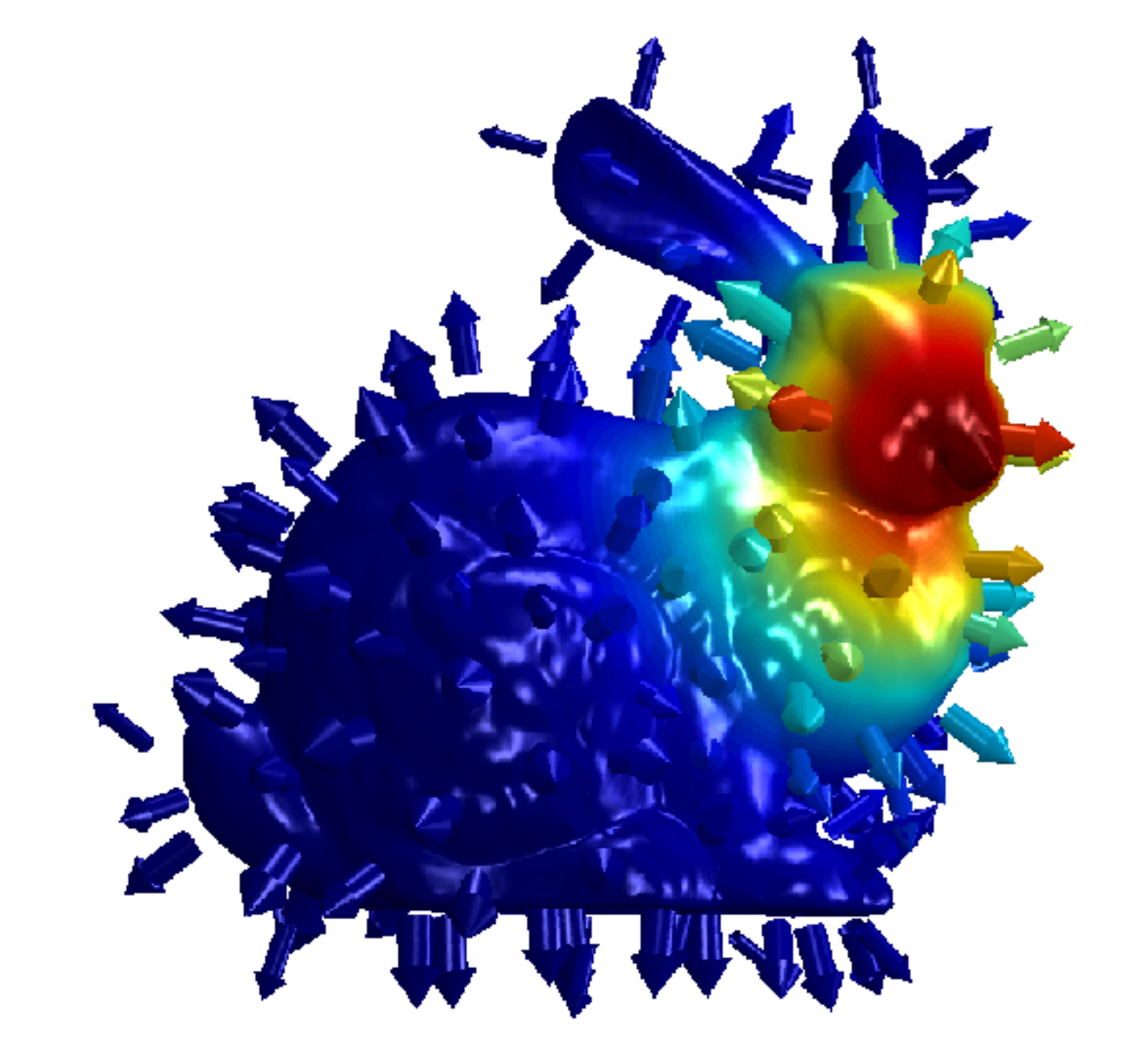} & \includegraphics[width=4.5cm, height=3.6cm]{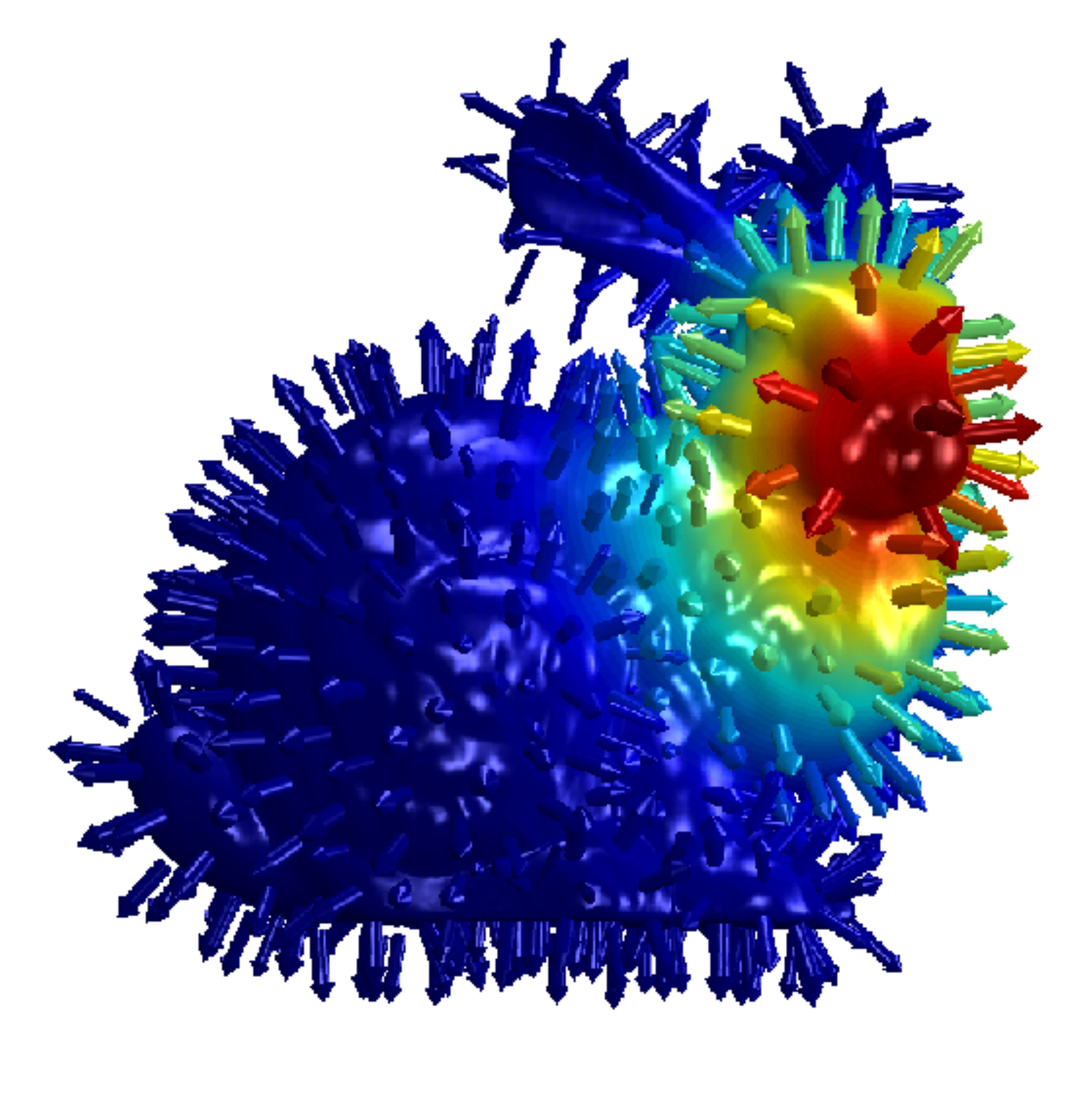} \\
$\lambda_{g}=0.04, \ \lambda_{f}=0.2, \ 57 \ Diracs$ & $\lambda_{g}=0.02, \ \lambda_{f}=0.1, \ 161 \ Diracs$ & $\lambda_{g}=0.01, \ \lambda_{f}=0.1, \ 571 \ Diracs$ 
\end{tabular}
\caption{Matching pursuit on a ``painted'' bunny with different parameters $\lambda_{g}$ and $\lambda_{f}$. Geometrically, the surface has $0.16 \times 0.22 \times 0.12$ extension in the 3D space and the signal goes from value zero (blue) to one (red). The original sampling of the fcurrent representation has 69451 Diracs and we choose a stopping criterion for the algorithm of $\epsilon = 5 \% $. The resulting Dirac fcurrents $\delta_{(x_{k},m_{k})}^{\xi_{k}}$ are here represented as colored vectors accordingly to the functional values $m_{k}$. Vectors are all of same length covering an area proportional to the norm of $\xi_{k}$. Notice that the sampling increases as $\lambda_{g}$ is smaller while the vector's colors are more accurate when $\lambda_{f}$ is smaller.}
\label{courantf_surface}
\end{figure}

\begin{figure}[!h]
\begin{tabular}{cc}
  \includegraphics[width=5.5cm, height=4cm]{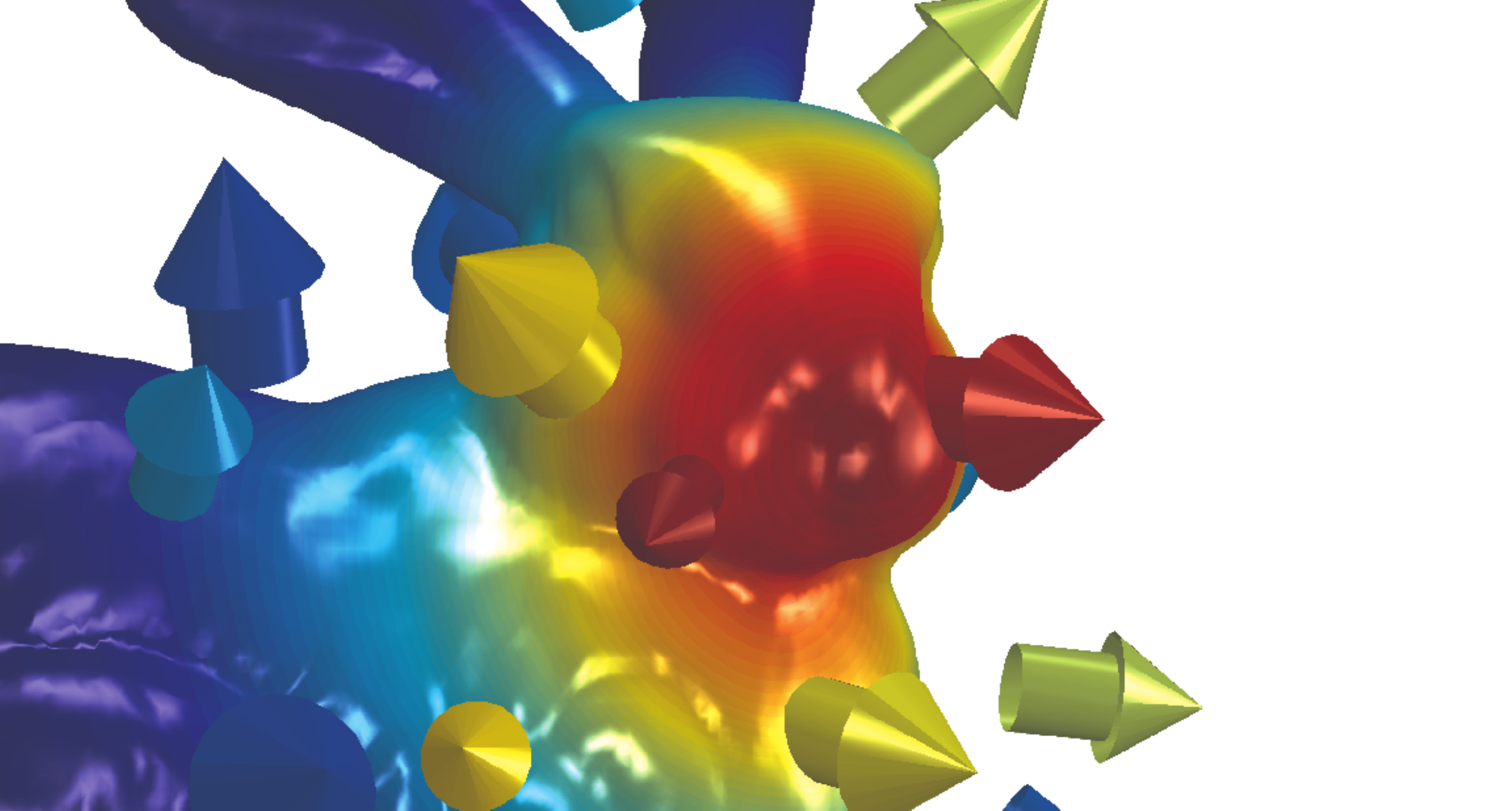} & \includegraphics[width=5.5cm, height=4cm]{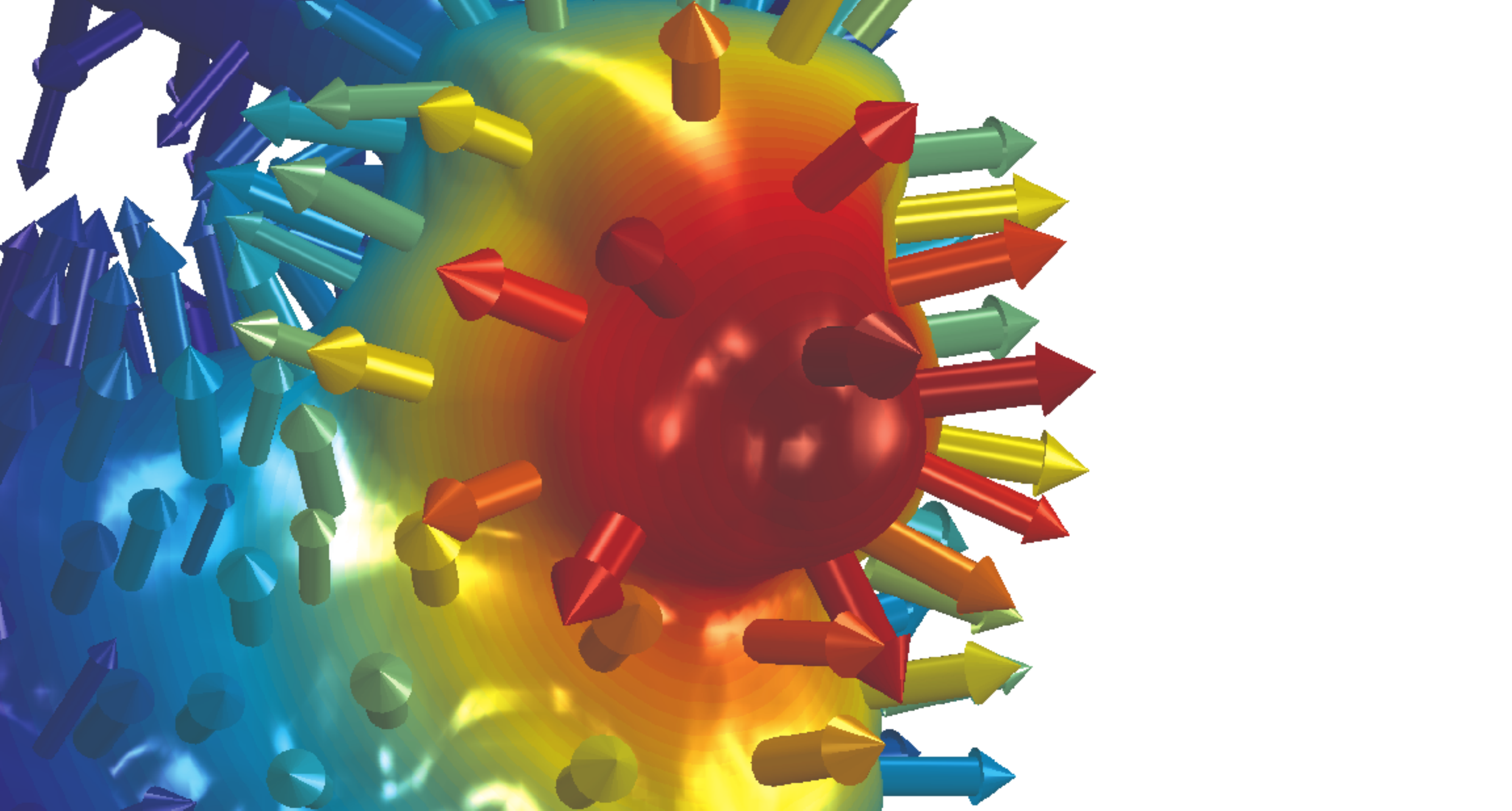} \\
$\lambda_{g}=0.04, \ \lambda_{f}=0.4$ & $\lambda_{g}=0.01, \ \lambda_{f}=0.1$ 
\end{tabular}
\caption{Close up on two of the previous results.}
\label{courantf_surface_closeup}
\end{figure}

\begin{figure}[!h]
\begin{center}
 \includegraphics[width=7cm]{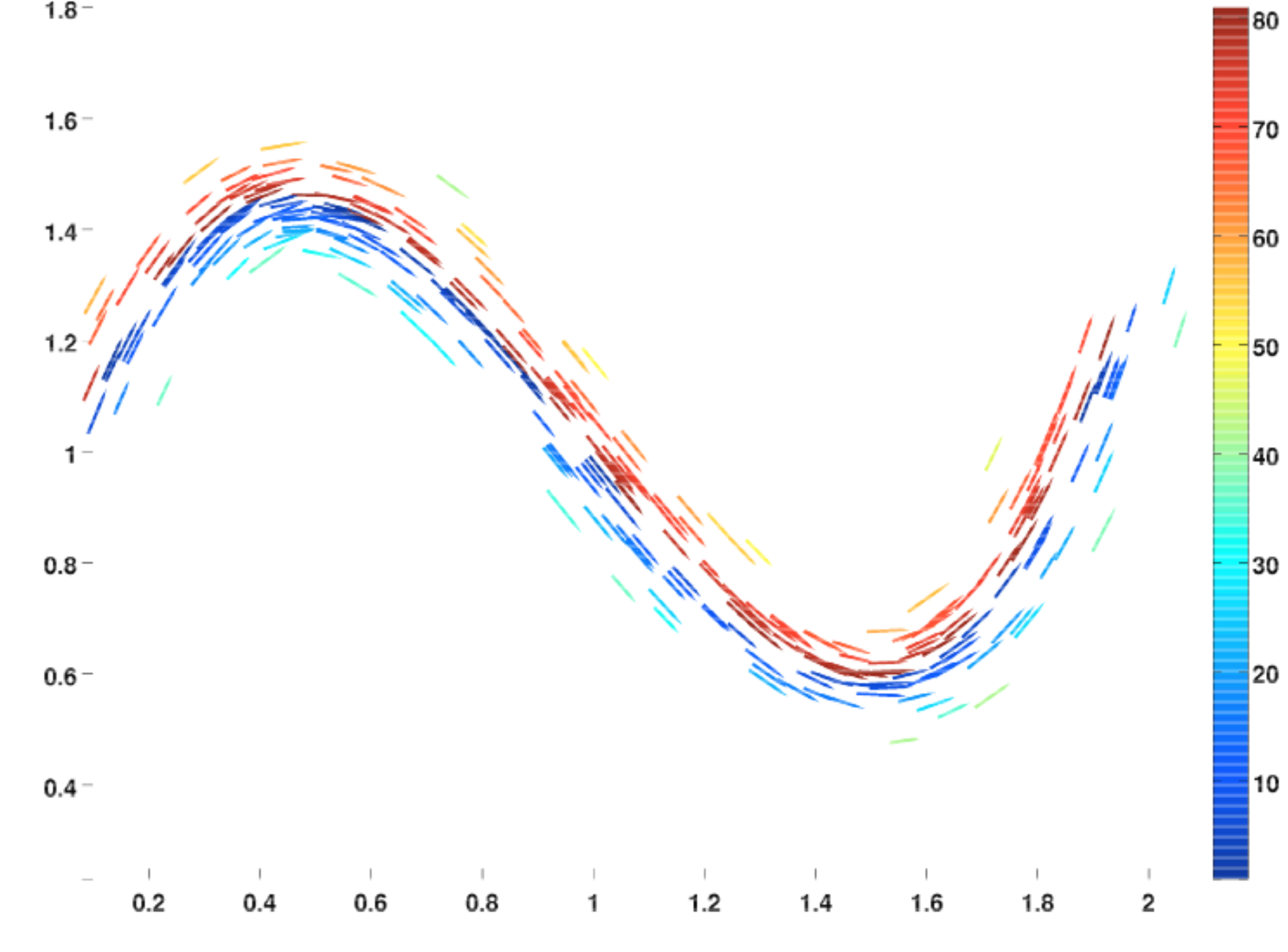}
\end{center}
\leftskip -1.5cm
\begin{tabular}{cc}
   \includegraphics[width=7cm]{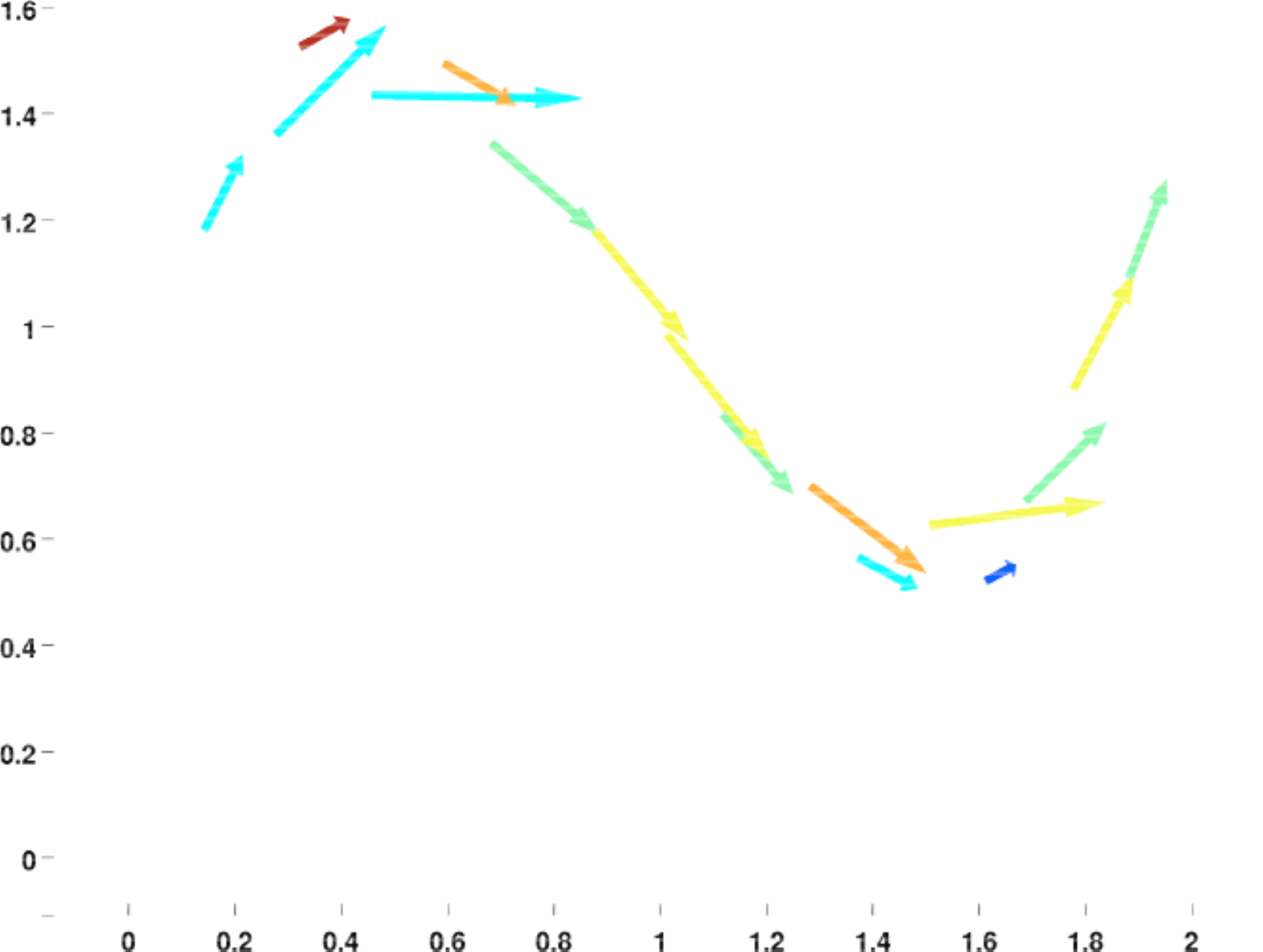} &
   \includegraphics[width=7cm]{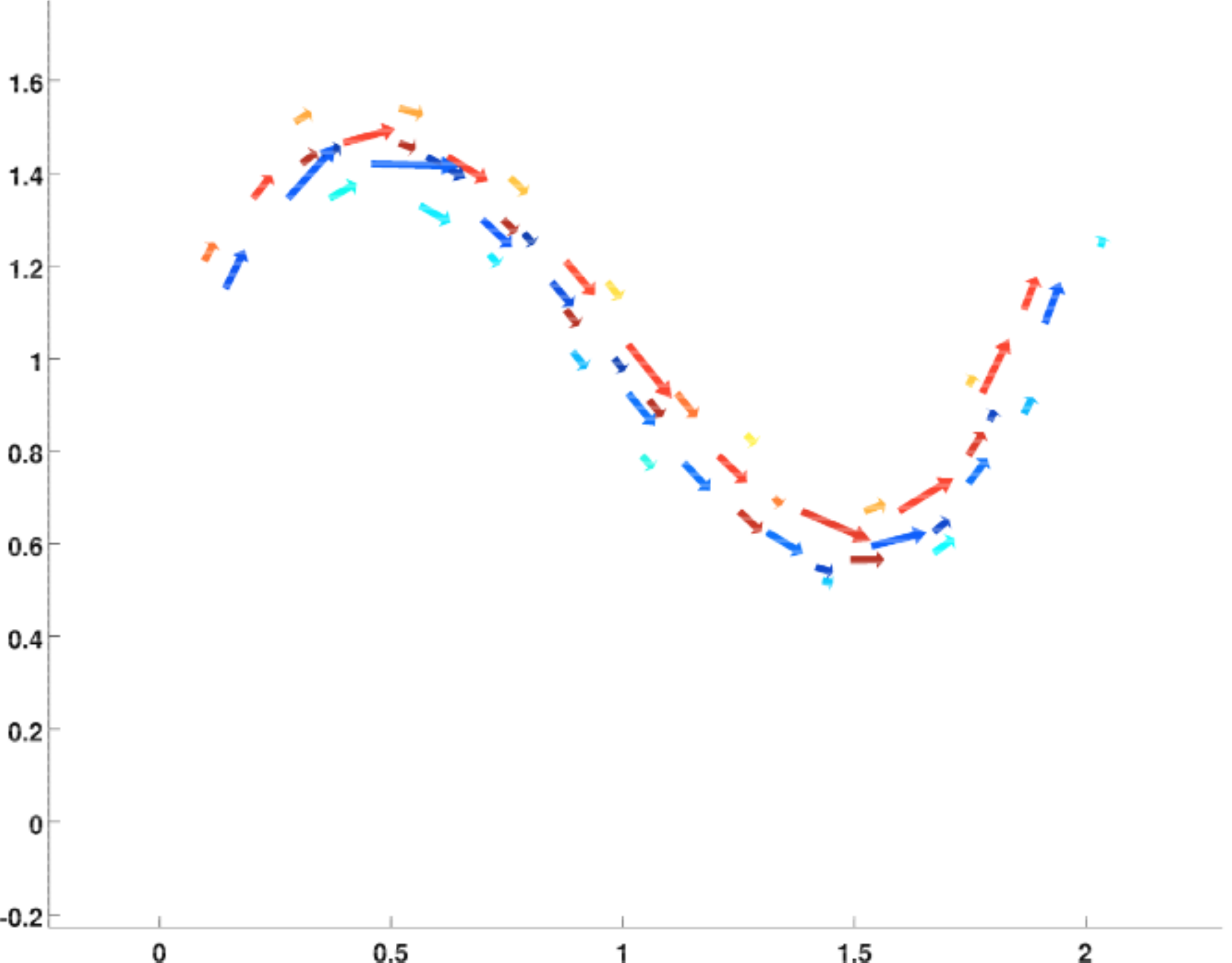} \\
\end{tabular}
\caption{Matching pursuit on a 2D fiber bundle, each fiber carrying one value of signal represented by the color. On top, the initial object consisting of 300 fibers. Below, we show two results of matching-pursuit with the same $\lambda_{g}$ but two different values for $\lambda_{f}$ : $\lambda_{f}=200$ for the left figure, $\lambda_{f}=20$ for the right one.}
\label{matching_pursuit_fibres}
\end{figure}

Here are now a few illustrative examples for real valued data on curves or surfaces. We will always consider kernels on fcurrents that are the tensor product of a Gaussian kernel in $\mathbb{R}^{3}$ of scale parameter $\lambda_{g}$ with a real Gaussian kernel in the signal space of scale parameter $\lambda_{f}$. In figure \ref{courantf_surface} and \ref{courantf_surface_closeup}, we emphasize the influence of both kernel sizes on the compression factor as well as on the precision of the functional values of the compressed shape. The bigger the parameter $\lambda_{g}$, the coarser the scale of representation is and fewer punctual fcurrents are therefore needed to compress shapes but more smaller features are lost. In figure \ref{matching_pursuit_fibres}, we focus more precisely on the compression's behaviour when computing matching-pursuit on a simulated fiber bundle of 2D curves carrying different signals. The scale $\lambda_{g}$ is the same for both figures but we show the results of matching-pursuit for two radically different values of $\lambda_{f}$. In both cases, matching pursuit provides an accurate approximation of the mean (accordingly to the kernel norm) with a very limited number of Diracs compared to the original sampling. However, note the important influence of $\lambda_{f}$. Taking a big value for this parameter means that the matching-pursuit will average values of the signals and provide a representation essentially with Dirac fcurrents having values for their signal parts close to the average (left figure) whereas for a smaller $\lambda_{f}$, the algorithm will only average the diracs that have close values of signal (right figure). 

In conclusion, these first examples of functional shape processing were meant to highlight that the combination of the fcurrent's representation with the use of RKHS metrics provides an easy solution to address the issue of redundancy and compression. The method provides important compression factors and enables \textit{scale analysis} on geometry and signal through the kernel parameters $\lambda_{g}$ and $\lambda_{f}$.

\newpage

\newpage

\subsection{A large deformation matching algorithm for functional shapes}
\label{subsec:match}
As a second illustrative example, we would like to briefly highlight the potentials of fcurrent representations in the context of computational anatomy and more generally in the context of shape spaces. It is clear that many important anatomical manifolds are coming with interesting data lying on it (for instance cortical thickness in anatomical MRI or activation maps in fMRI scans among many possibilities) and are perfect examples of functional shapes as defined in this paper. The statistical analysis of a population of such functional shapes is however a real challenge since the relevant information in a functional shape may be buried in two sources~: the pure geometrical shape defined by the manifold itself and the signal information spread on the support. However the geometrical and functional parts are more likely intertwined with each other. 

When only pure geometrical shapes are considered, the concept of shape space equipped with a Riemannian metric offers proper tools for the local analysis of a population of shapes seen as a distribution of points in a shape space. In particular, the use of Riemannian exponential map around a template conveys an efficient linearization of the shape space to describe the differences between shapes. However, observed shapes are contaminated by many errors coming from various pre-processing pipelines driving the extraction of shapes from raw data and the shape space is not sufficient to accommodate any observed shape. Moreover, and more fundamentally, shapes in a shape space are \emph{ideal} exemplars of real shapes with controlled complexity to address properly estimation issues from a limited sample. Consequently a discrepancy measure or a noise model is needed to link ideal shapes in shape space with observed shapes. A coherent framework is provided by the \emph{current} framework~: indeed observed shapes can be represented as a vector in a Hilbert space of currents which is also embedding a Riemannian shape space $\mathcal M$ of ideal shapes~: $\mathcal M \hookrightarrow {W'}$ so that a population of observed shapes $(S_i)$ can be represented as a sum $S_i=m_i+r_i$ where $m_i\in\mathcal M$ and the residual noise $r_i\in {W'}$.
Introducing a template $m_0$ and using the linearization provided around $m_0$ by the Riemannian exponential map $\text{Exp}_{m_0}$ we can write for any observed shape $S$:
\begin{equation}
S=\text{Exp}_{m_0}(u)+r\label{eq:u,r}
\end{equation}
where $(u,r)\in T_{m_0}\mathcal M\times {W'}$. Note that the $(u,r)$ are lying in a vector spaces and $t\mapsto m_t\doteq \text{Exp}_{m_0}(tu)$ is a geodesic on $\mathcal M$. Introducing the metric $\|\ \|_{m_0}$ at $m_0$ and the metric $\|\ \|_{W'}$ on ${W'}$, we can estimate an optimal decomposition (\ref{eq:u,r}) $(u(S),r(S))$ of an observed shape $S$ by the minimization of $\|u\|_{m_0}^2+\|r\|_{W'}^2$. 

When pure geometrical shapes are no longer involved but functional shapes instead, the previous setting breaks down with usual currents but is still valid if ${W'}$ is replaced by a RKHS space of \emph{fcurrents}. The space $\mathcal M$ itself can be defined as $\mathcal M=\{\ g\cdot m_0\ |\ g\in G\}$ i.e. the orbit of a template $m_0$ under the action of a group of deformations $G$. The diffeomorphic transport discussed in subsection \ref{subsec:diff_transp} offers several examples of such action. We will consider the simple situation of functional shapes with real valued signals ($E=\mathbb{R}^d,\ M=\mathbb{R}$) where the action is given by (\ref{eq:trivial_action}) even if more complex actions as defined by (\ref{eq:phi,psi}) and (\ref{eq:phi,psi,dual}) could be used. In this setting, the Riemannian structure on $\mathcal M$ is inherited from the optimisation of the kinetic energy $\int_0^1\|v_t\|_V^2dt$ on a time dependant Eulerian velocity fields $(t,x)\mapsto v(t,x)$ of the trajectory $t\mapsto \phi_t\cdot m_0$  where $\phi_t$ is the flow of the ODE $y'=v(t,y)$ starting from the identity. The overall framework has been popularized as the large deformation diffeomorphic mapping setting (LDDMM). The space $V$ is a RKHS space of vector fields, here given by an isotropic Gaussian kernel, generating a right invariant distance on the group $G$ of diffeomorphisms generated by flows of kinetic energy. This induces, by Riemannian submersion, a Riemannian structure on $\mathcal M$ (see \cite{Miller2002,Trouve2005a} for a more extended presentation of this geometrical setting). In particular, if $m_0=C_{(X,f)}$ with $X$ is a smooth manifold with finite volume or if $m_0=C$ is a more general element of $W'$ such that $M(C)<\infty$ (for instance a countable family of $(X_i,f_i)$'s with $\sum \text{vol}(X_i)<\infty$) then the continuity result given by Proposition \ref{propcontrolnormdefor} or (\ref{eq:cont_action}) gives the continuous embedding  $\mathcal M\hookrightarrow W'$.

Obviously the RKHS norm plays the role of an attachment term and could be coupled with other matching approaches (even if we think that the previous setting is particularly attractive for further statistical studies). The reader not familiar with the above geodesic setting could replace the mapping $u\mapsto m_1(u)=\text{Exp}_{m_0}(u)$ by any other mapping $u\mapsto m_1(u)$.

\begin{figure}[htb]
\begin{tabular}{cc}
   \includegraphics[width=6cm,height=5cm]{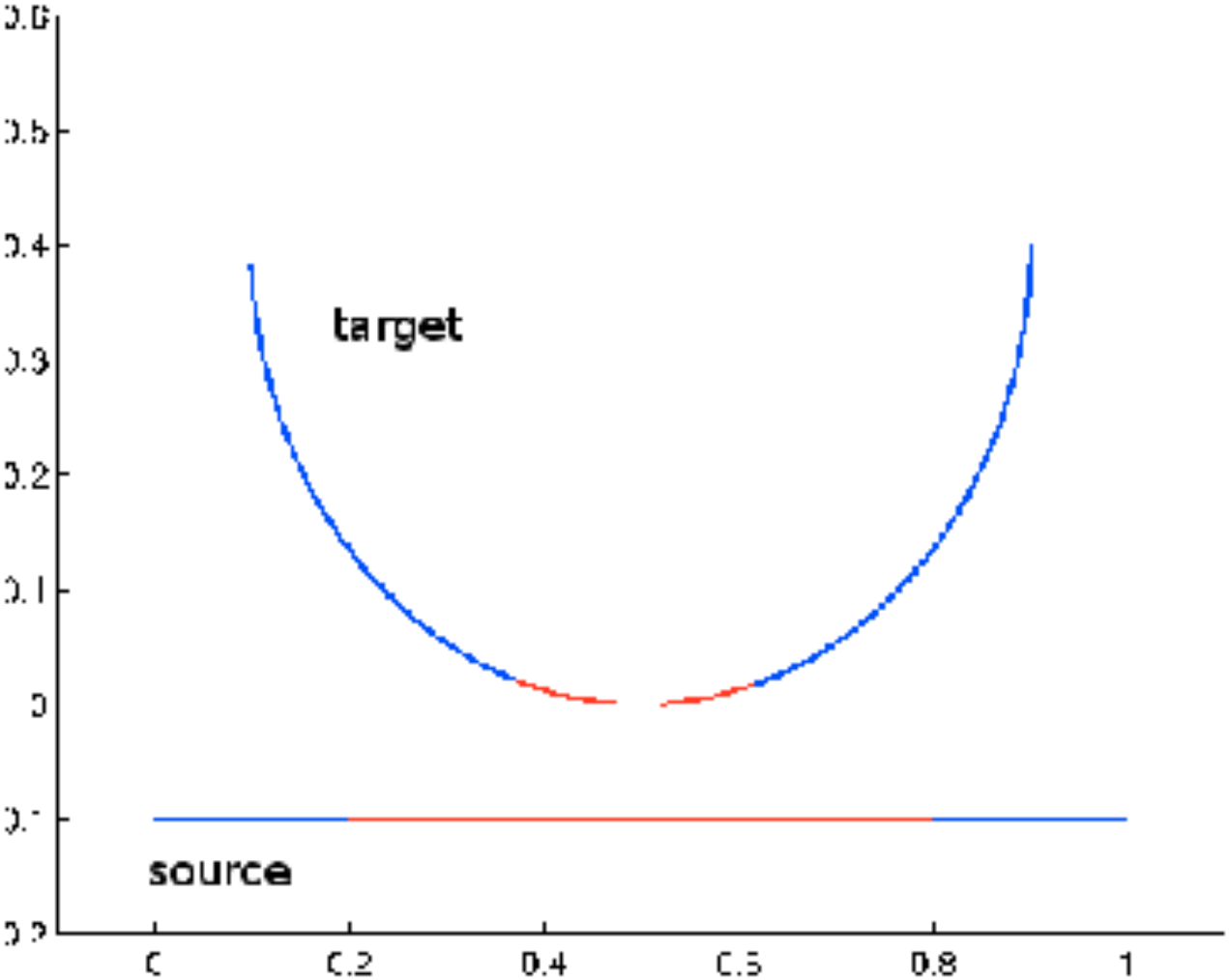}  & \includegraphics[width=6cm,height=5cm]{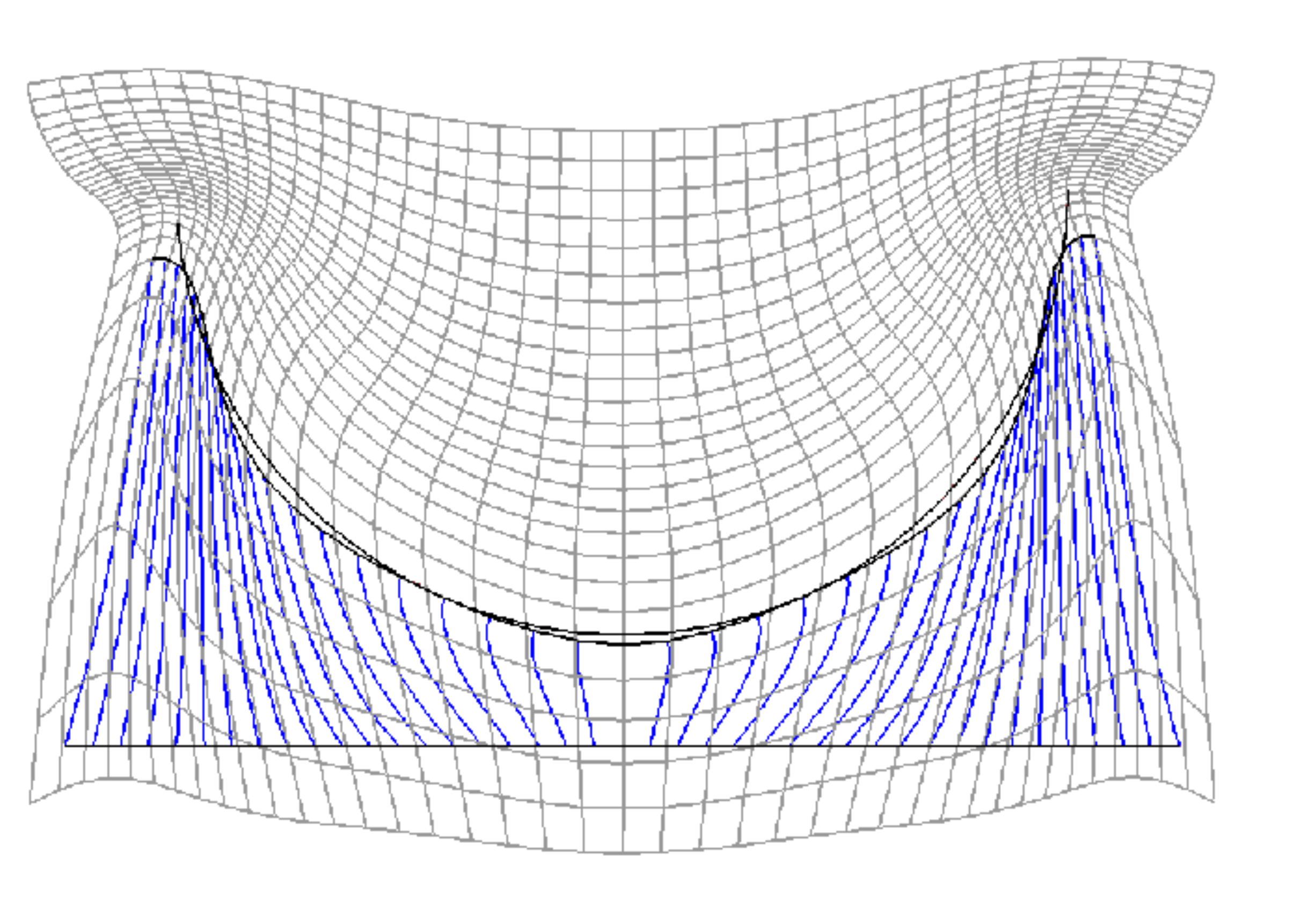} \\
   \includegraphics[width=6cm,height=5cm]{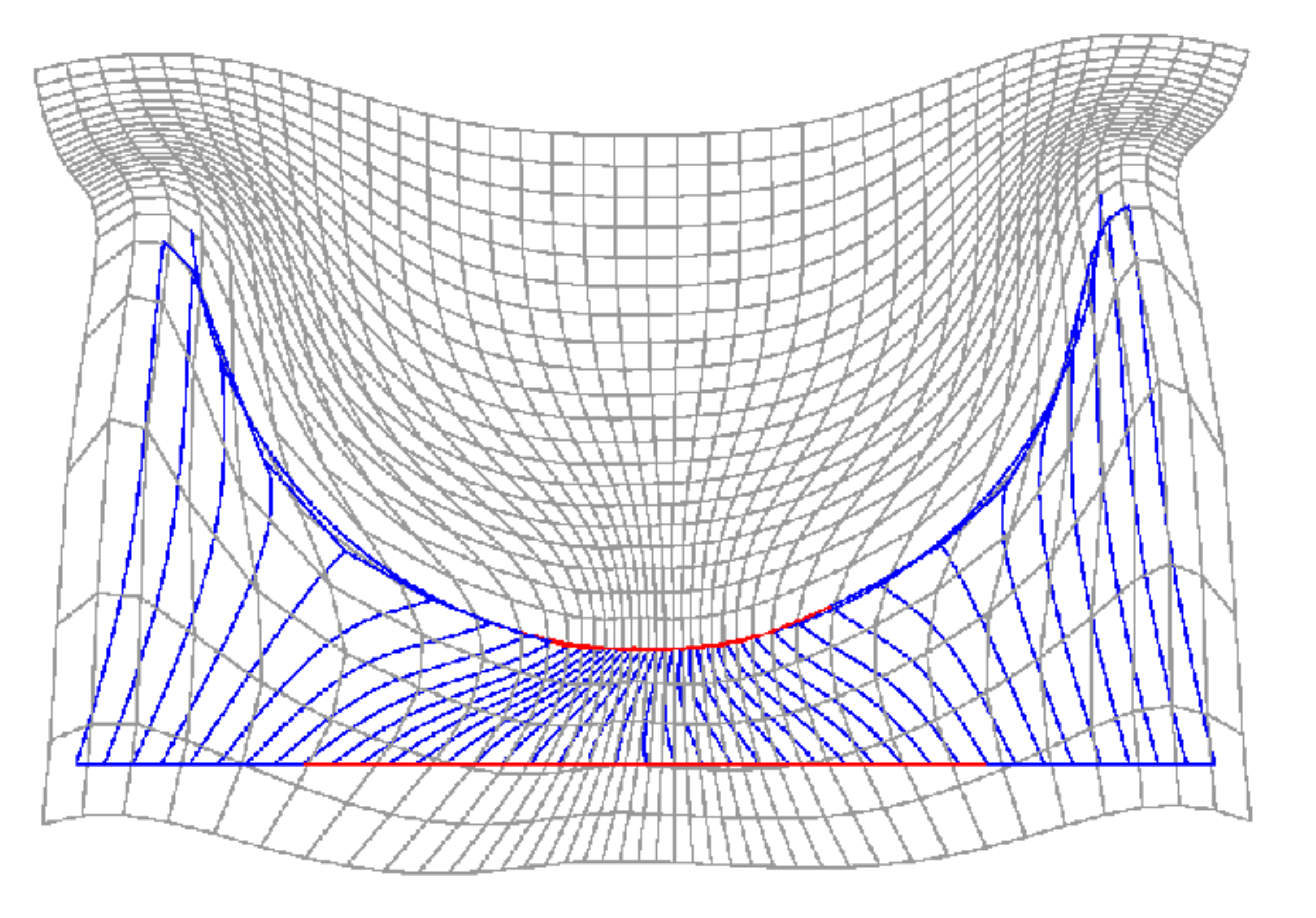} & \includegraphics[width=6cm,height=5cm]{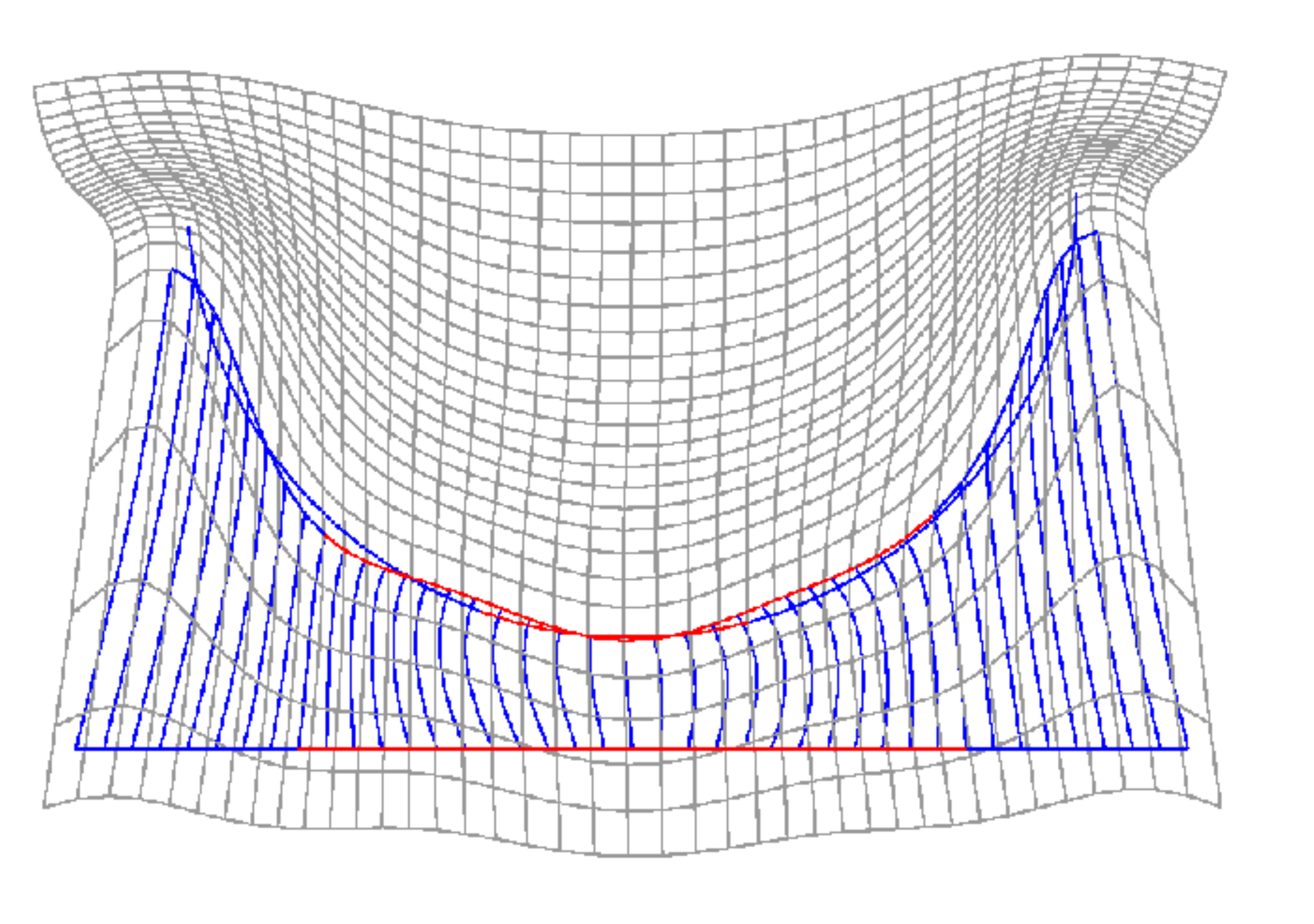} 
\end{tabular}
\caption{Example of registration of two functional curves (top left) with binary signal (blue is zero and red is one). On top right, we show the classical matching with currents on the purely geometrical curves. On bottom left, the same curves are matched with our extension of LDDMM to functional currents. In both cases, the deformed curve fits closely to the target one but note the difference of the deformation field for the functional current's approach. Finally, on the right, we show the result of matching we obtain again with fcurrents' LDDMM but with a big value of $\lambda_{f}$ compared to the signal, in which case the matching is nearly similar to the current matching.}
\label{matching_grid}
\end{figure}

\begin{figure}[!h]
\leftskip -2cm
\begin{tabular}{cc}
   \includegraphics[width=7cm,height=6cm]{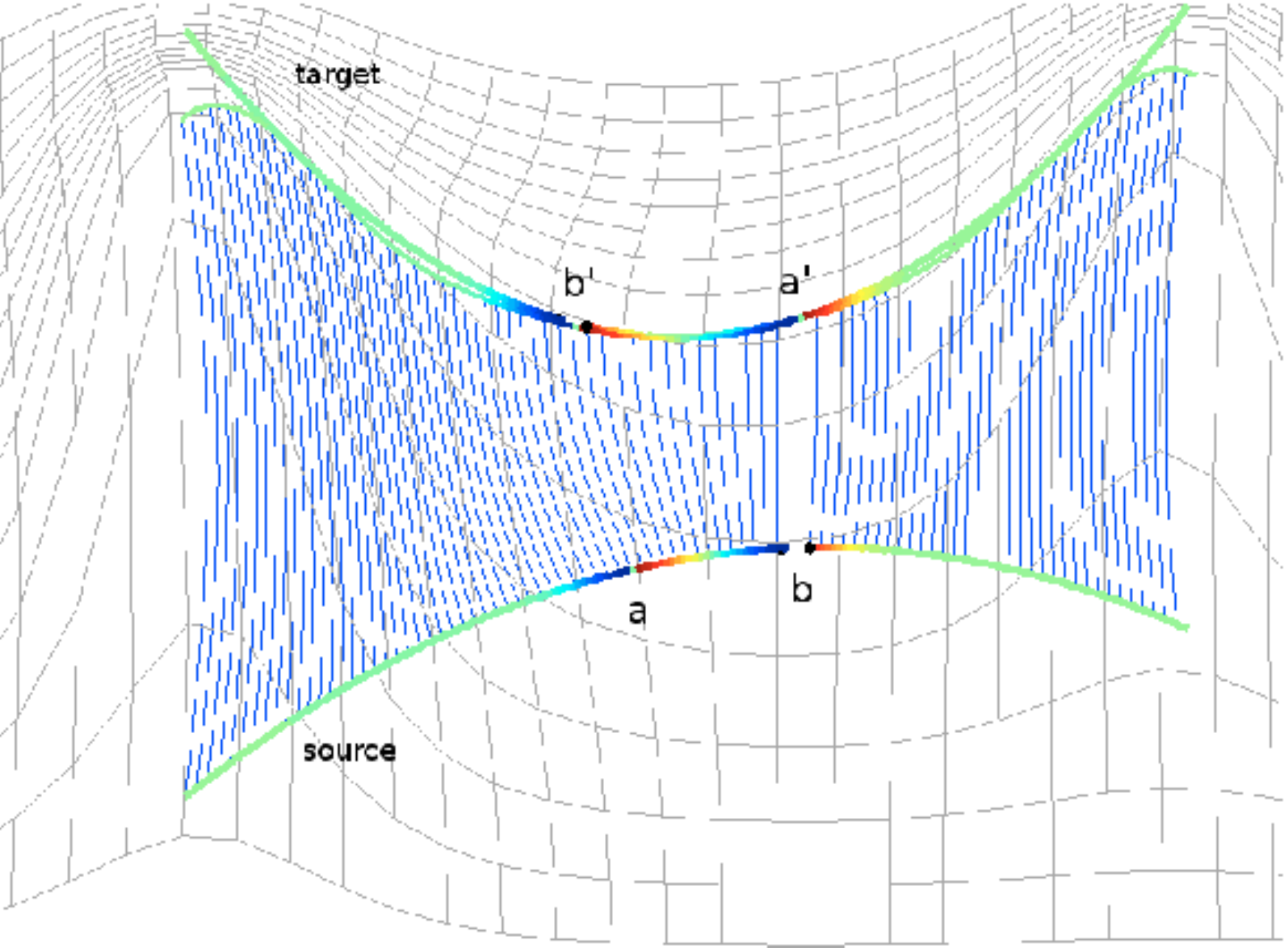} &
   \includegraphics[width=7cm,height=6cm]{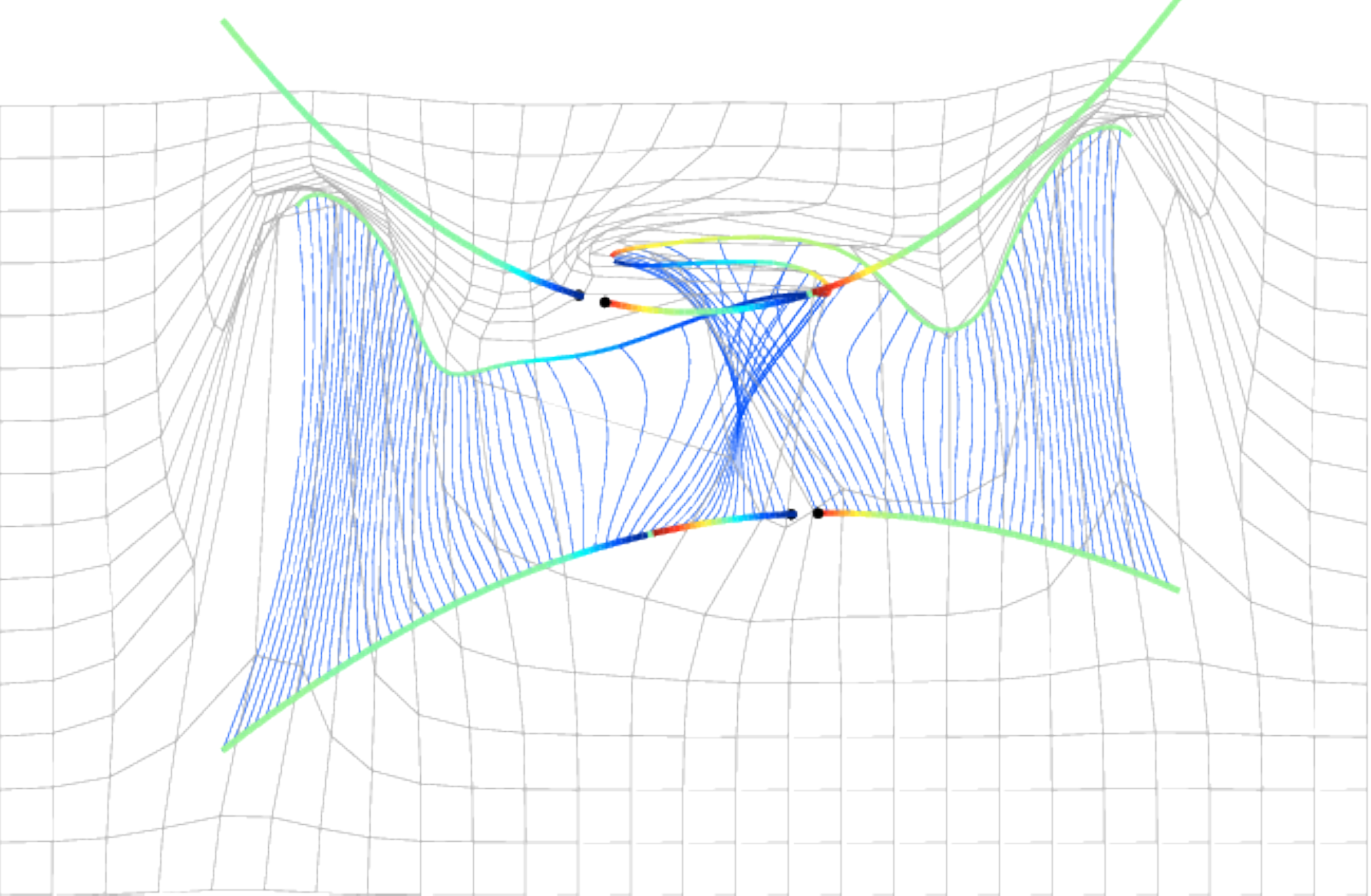} 
\end{tabular}
\caption{LDDMM matching of two planar curves with discontinuous signals and topological disconnections. Each curve has two points of functional discontinuity, one of them being also a disconnection of the geometrical support (point b on the source and b' on the target). On the right figure, the matching is performed by representing the colored curve as a current in the product space $\mathbb{R}^{2}\times \mathbb{R}$ as explained in section \ref{subsec:func}. On the left, with the functional currents' representation. We see that the resulting deformation is much perturbed by the disconnections in the case of product currents : the algorithm intends to match connected part of the source shape on a connected part of the target shape although it leads to a very unnatural matching.}
\label{matching_morceaux}
\end{figure}

\begin{figure}[htb]
\leftskip -1.5cm
\begin{tabular}{ccc}
   \includegraphics[width=4cm,height=6cm]{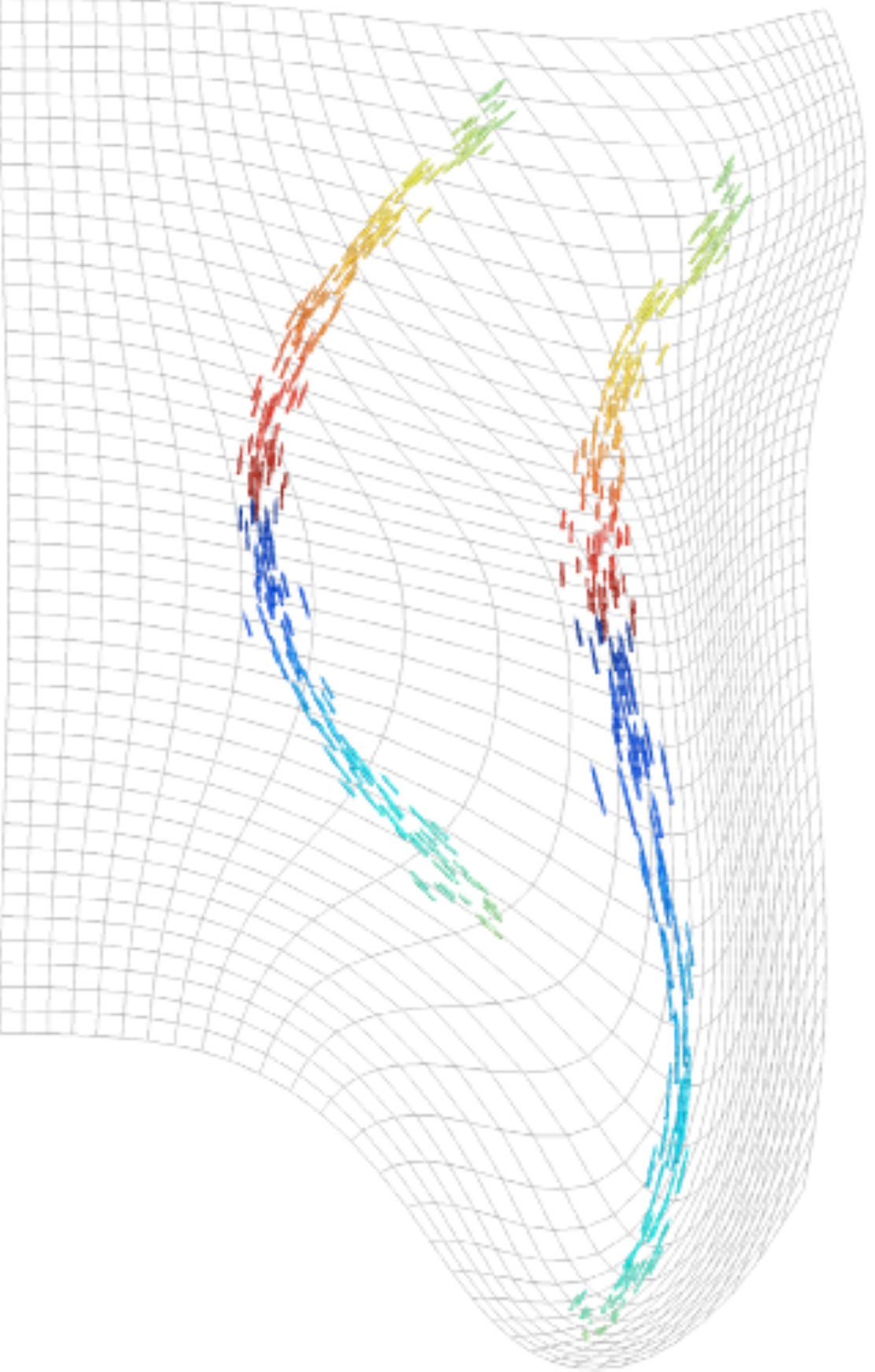} &
   \includegraphics[width=6cm,height=6cm]{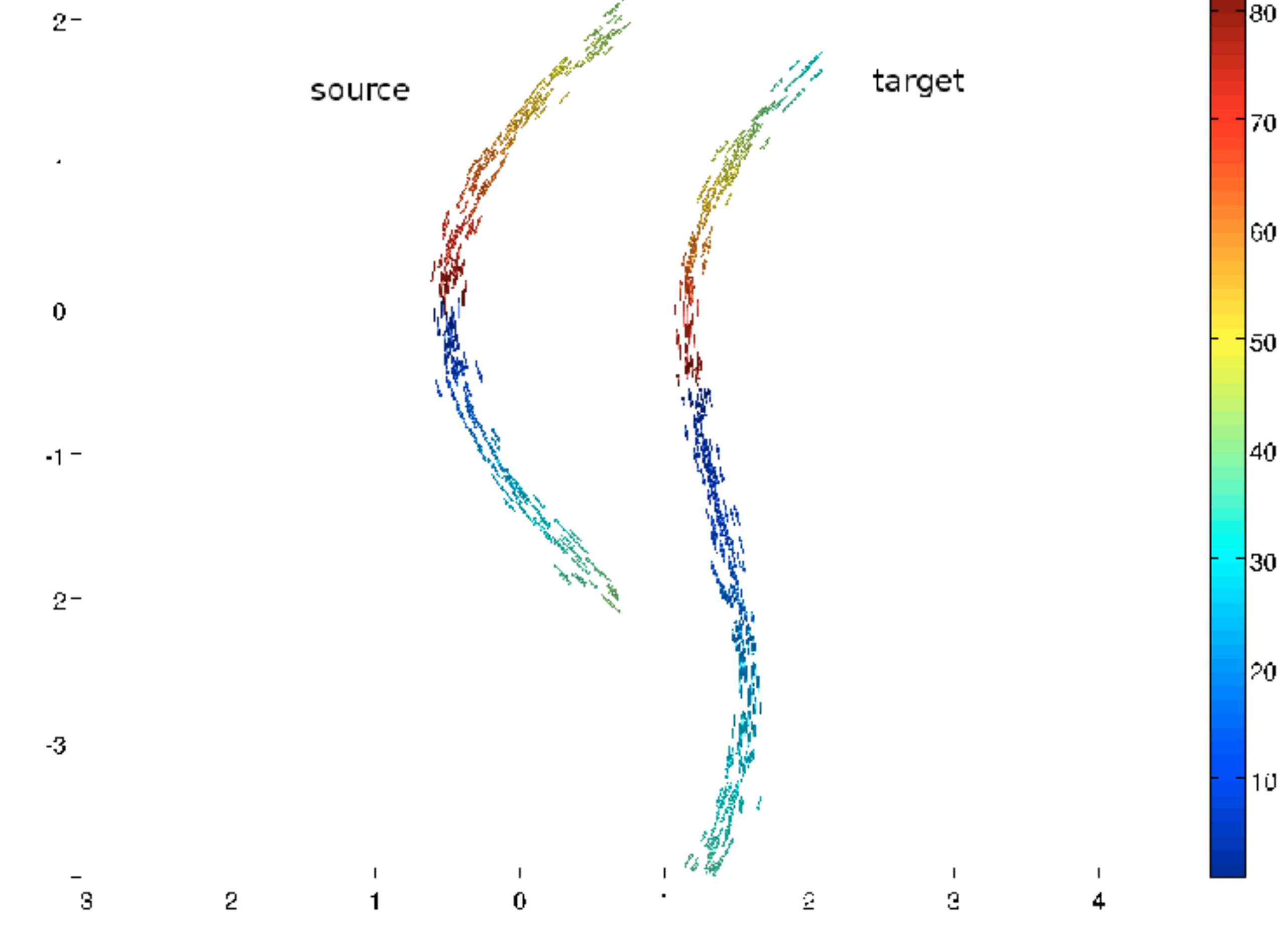} &
   \includegraphics[width=4cm,height=6cm]{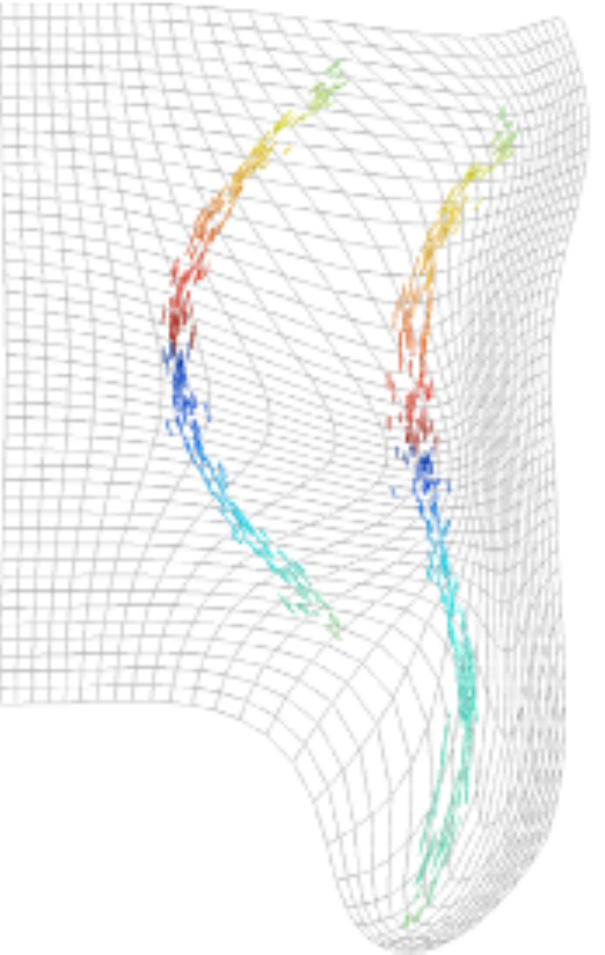} \\
   Functional current matching & Source and target & Current matching
\end{tabular}
\caption{A last example of matching on the case of a fiber bundle with signal. On the center figure, the source and target functional shapes. On the left, the resulting matching with the deformed shape and the deformation grid for the functional currents' setting. On the right, the result obtained by matching with currents. Note that even if the geometrical shapes are well matched in both cases, the two deformations are not the same. Functional currents elongate the dark blue part to fit with the target shape's colors whereas currents, by not taking signal into account, shrinks it.}
\label{matching_fiberbundle}
\end{figure}

With attachment distances provided by the RKHS norms on fcurrents, it is then possible to extend LDDMM algorithm to the registration of functional shapes. Leaving the technical details of implementation to a future paper, we just present some results of the method on simple examples. As we can expect, the resulting matching is driven both by the geometry of the shapes and by the functional values they carry accordingly to the scales of both kernels, which we first show on the example of figure \ref{matching_grid}. 
If we compare it now to the colored currents of section \ref{subsec:func}, we see that since functional currents clearly separate signal and geometry, we no longer have the same drawbacks : in the colored surfaces of figure \ref{courantscolor}, we have shown on the right the matching result with the functional currents' approach. In addition, the functional current representation is totally robust both to punctual outlying signal values and to missing connections between points, which is clear from the definition of the RKHS norm, because geometrically negligible subsets of the shape have zero norm. It was not the case for instance with the product current idea (cf \ref{subsec:func}) since variations of signal also carry non-zero norm. This has important consequences when trying to match curves with missing connections as we show on the example of figure \ref{matching_morceaux}. In our sense, it makes functional currents more fitted to the treatment of fiber bundles carrying signal, like the example given in figure \ref{matching_fiberbundle}. 

A second important thing to point out is that having a norm defined by the tensor product of two kernels $K_{g}$ and $K_{f}$ with two independent scales provides a total flexibility for the matching, geometrically and functionally. The choice of a bigger parameter $\lambda_{f}$ for instance allows the matching of signal values to be accurate only at a bigger scale, hence our method could still achieve matching under noisy or imprecise signals on shapes. The counterpart is of course the presence of an additional parameter that must be adapted to the data, based upon an a priori on the reliability of the signals we want to match. Multi-scale approaches can also be built by adding kernel at different scales in the spirit of \cite{Risser2011} or \cite{Sommer2011}. But still, functional currents encompass usual currents' approach in the sense that for the limit case $\lambda_{f}\rightarrow \infty$, matching with fcurrents will reduce to a classical matching of purely geometrical parts of the data (cf bottom left figure \ref{matching_grid}). 

\newpage

\section{Conclusion and outlook}
We have presented in this paper a way to formally generalize the notion of currents in the purpose of integrating functional shapes into a coherent and robust representation. Functional currents provide a framework to model geometrically-supported signals of nearly any nature and regularity while preserving the interest of the current's approach for computational anatomy. The second main point of the study is the definition of an appropriate norm. The definition of a RKHS structure provides a distance between functional shapes that enjoys worthy control properties as stated in section \ref{subsec:control}. At the same time, the resulting Hilbert structure on fcurrents opens the way to a very wide class of applications. Although numerical issues that appear when computing with currents were not detailed in this paper, we have presented two examples of processing algorithms for functional shapes~: a matching pursuit scheme to address fcurrents' compression and averaging as well as an adaptation of LDDMM algorithm for diffeomorphic registration of two functional shapes. Examples were provided essentially in the simplest cases of curves or surfaces with real-valued signal but same methods could easily apply to different kind of manifold, signal and deformation models.\\
To sum up, the article has essentially the objective of setting a path to extend the scope of traditional computational anatomy to these kind of data structures we called functional shapes. This might constitute a serious possibility to improve registration and statistical estimation of deformable templates, which constitutes the future step of our work. In the case of brain anatomy for instance, by taking into account the additional information on the cortical surfaces provided by fMRI maps or estimations of cortical thickness. And last but not least, let us insist again on the point that the RKHS distances we have derived between functional shapes enable joint comparison of geometry and function without the usual curse of requiring a point to point correspondence between shapes or common coordinate systems, which opens interesting possibilities with respect to statistics on functional shapes. 

\addcontentsline{toc}{chapter}{Appendix}
\addtocontents{toc}{\protect\setcounter{tocdepth}{-1}}

\section*{Acknowledgments}
This work was made possible thanks to HM-TC (Hippocampus, Memory and Temporal Consciousness) grant from the ANR (Agence Nationale de la Recherche).

\appendix
\section{Deformations' modelling in the LDDMM framework}
In this appendix, we remind a few intermediate results which are necessary for the full proof of proposition \ref{propcontrolnormdefor}. Most of them refer to deformations' modelling and can be found either in \cite{Glaunes} or \cite{Younes} (chap. 12).  \\
\\
Using notations of \cite{Younes}, for $p \in \mathbb{N}$, let $C_{0}^{p}(\mathbb{R}^{n},\mathbb{R}^{n})$ be the Banach space of $p$-times continuously differentiable vector fields $v$ on $\mathbb{R}^{n}$ such that $v,dv,..,d^{p}v$ vanish at infinity, which is equipped with the norm $|v|_{p,\infty}= \sum_{i=1}^{p} |d^{i}v|_{\infty}$. \\
Now, let $\chi^{p}$ be the set of integrable function from the segment $[0,1]$ into $C_{0}^{p}(\mathbb{R}^{n},\mathbb{R}^{n})$. Any element of $\chi^{p}$ is a time-varying vector field we will denote $v(t,.)$, $t \in [0,1]$. On $\chi^{p}$ we define the norm :
\begin{equation*}
 \| v \|_{\chi^{p}} = \int_{0}^{1} |v(t,.)|_{p,\infty} dt
\end{equation*}
Note that we have $\chi^{p} \subset \chi^{p-1} \subset...\subset \chi^{0}$ and that if $v \in \chi^{p}$, $\| v \|_{\chi^{0}} \leq... \leq \| v \|_{\chi^{p}}$. \\
For any $v \in \chi^{1}$, we consider the differential equation $\frac{dy}{dt} = v(t,y)$ with initial condition $y(s)=x \in \mathbb{R}^{n}$ at time $s \in [0,1[$. We have :
\begin{theo}
 For all $x \in \mathbb{R}^{n}$ and $s \in [0,1[$, there exists a unique solution on $[0,1]$ of the differential equation $\frac{dy}{dt} = v(t,y)$ such that $y(s)=x$. We denote by $\phi_{s,t}^{v}(x)$ the value at time $t$ of this solution. $(t,x)\mapsto \phi_{s,t}^{v}(x)$ defined on $[0,1] \times \mathbb{R}^{n}$ is called the \textbf{flow} of the differential equation.
\end{theo}
In other words, the flow satisfies the following integral equation :
\begin{equation}
\label{eq_flot}
 \phi_{s,t}^v(x) = x + \int_{s}^{t} v(r,\phi_{s,r}^{v}(x)) dr
\end{equation}
We then have :
\begin{theo}
 For all $v \in \chi^{1}$ and all $s,t \in [0,1]$, $\phi_{s,t}^{v}$ is a $C^{1}$-diffeomorphism of $\mathbb{R}^{n}$. In the special case where $v=0$, $\phi_{s,t}^{v}$ is the identity application.
\end{theo}

From equation \ref{eq_flot}, using Gronwall inequality, it is easy to show that :
%
\begin{theo}
\label{cor_diffeo}
For all $R>0$ there is a constant $C(R) > 0$ such that, for all $v \in \chi^{1}$ with $\|v\|_{\chi^{1}}\leqslant R$ :
\begin{equation*}
 \| \phi_{s,t}^{v} - Id \|_{\infty} \leqslant C(R).\|v\|_{\chi^{0}} \leqslant C(R).\|v\|_{\chi^{1}}
\end{equation*}
\end{theo}

In a similar way, the same kind of control can be obtained for the differential of the flow as stated below :
\begin{theo}
For all $v \in \chi^{1}$ and $s,t \in [0,1]$, $\phi_{s,t}^{v}$ is a $C^{1}$ function whose differential satisfies the integral equation :
\begin{equation*}
 d_{x}\phi_{s,t}^v = Id + \int_{s}^{t} d_{x} v(r,\phi_{s,r}^{v}(x)) dr
\end{equation*}
\end{theo}

\begin{theo}
For all $R>0$ there is a constant $C(R) > 0$ such that, for all $v \in \chi^{1}$ with $\|v\|_{\chi^{1}}\leqslant R$ :
\begin{equation*}
 \| d\phi_{s,t}^{v} - Id \|_{\infty} \leqslant C(R).\|v\|_{\chi^{1}}
\end{equation*}
\end{theo}

This last result, together with the multilinearity of exterior product and jacobian, leads to the following corollary :
\begin{cor}
\label{cor_jac_dvect}
For $\|v\|_{\chi^{1}}$ small enough, there exists constants $\alpha>0$ and $\beta>0$ such that for all $x \in E$ : 
\begin{eqnarray*}
| Jac_{x}(\phi_{s,t}^{v}) - 1 | &\leq& \alpha \|v\|_{\chi^{1}} \\
\| d_{x}\phi_{s,t}^{v}(\xi_{1})\wedge...\wedge d_{x}\phi_{s,t}^{v}(\xi_{d}) - \xi_{1} \wedge ... \wedge \xi_{d} \| &\leq& \beta \|v\|_{\chi^{1}}. \|\xi_{1}\wedge ... \wedge \xi_{d} \|
\end{eqnarray*}
\end{cor}

\newpage 

\bibliographystyle{abbrv}
\bibliography{biblio2,LDDMM}
\end{document}